\documentclass[a4paper,11pt,english]{article}
\usepackage{a4wide}
\usepackage{babel,csquotes}
\usepackage{amssymb,amsmath}
\usepackage[small,justification=justified,singlelinecheck=false]{caption}
\usepackage{tikz}
\usepackage{float}
\usepackage{tensor}
\usepackage{longtable}
\usepackage{tabu}
\usepackage{colortbl}
\usepackage[titles]{tocloft}
\usepackage{hyperref}
\usepackage{pbox}
\usepackage{ytableau}
\usepackage{textcomp}
\usepackage{makerobust}
\usepackage[
	backref=true,
	sorting=none,
	maxnames=99,
	firstinits=true,
	doi=true,
	citestyle=numeric-comp,
	backend=biber
	]{biblatex}

\usepackage[T1]{fontenc}
\usepackage{lmodern}

\addto\extrasenglish{%
}
\addto\extrasenglish{%
}

\makeatletter 
\def\leqn{\tagsleft@true} 
\def\reqn{\tagsleft@false} 
\def\fleq{\@fleqntrue\let\mathindent\@mathmargin \@mathmargin=0pt} 
\def\cneq{\@fleqnfalse} 
\makeatother

\newenvironment{mma}{%
\fleq%
\tabulinesep = 4pt%
\begin{equation}%
\begin{tabu} to 0.93\linewidth {@{\;\;\;}r@{\;\;}X@{}}%
}{%
\end{tabu}%
\end{equation}%
\cneq%
}%

\newenvironment{mma*}{%
\fleq%
\tabulinesep = 4pt%
\begin{equation*}%
\begin{tabu} to 0.93\linewidth {@{\;\;\;}r@{\;\;}X@{}}%
}{%
\end{tabu}%
\end{equation*}%
\cneq%
}%

\newcommand{\mmain}[1]{\mmaintext &  \texttt{\bfseries #1}}
\newcommand{\mmanoin}[1]{\phantom{\mmaouttext} & \texttt{\bfseries #1}}
\newcommand{\mmaout}[1]{\mmaouttext & $\tt{#1}$ }
\newcommand{\mmanoout}[1]{\phantom{\mmaouttext} & $\tt{#1}$ }

\newcommand{\mmaouttext}{\mmainouttext{Out\,=}}
\newcommand{\mmaintext}{\mmainouttext{\;\;In\,:=}}
\newcommand{\mmainouttext}[1]{\textrm{\normalfont{\small\textcolor{darkgray}{#1}}}}

\newcommand{\ruledelayed}{\tikz[baseline=-\the\dimexpr\fontdimen22\textfont2\relax]{\filldraw (0.04,0.05) circle (0.6pt);	\filldraw (0.04,-0.05) circle (0.6pt); \draw[line width=0.8pt] (0.07,0) -- (0.25,0); \draw[line width=0.8pt] (0.16,0.08) -- (0.25,0) -- (0.16,-0.08);}\;}
\newcommand{\uunderline}[1]{\underline{\underline{#1}}}

\newcommand{\xAct}{\emph{xAct}}
\newcommand{\xTras}{\emph{xTras}}

\newcommand{\func}[1]{{\ttfamily #1}}
\newcommand{\funcref}[1]{\hyperref[#1]{{\ttfamily\nameref*{#1}}}}

\newcommand{\tab}{\hspace*{5mm}}

\newcommand{\usageline}[2]{%
	&&& \\[-6pt]
	& \multicolumn{2}{l}{{\ttfamily #1}} & \\
	& & #2 & \\[4pt]
	\hline
}
\newcommand{\usagelinelong}[2]{%
	&&& \\[-6pt]
	& \multicolumn{2}{l}{{\ttfamily #1}} & \\
	& & #2 & \\[10pt]
	\hline
}

\newsavebox{\coloredbgbox}
\newenvironment{usagetable}
  {\begin{lrbox}{\coloredbgbox}\begin{tabu}{llXl	} \hline}
  {\end{tabu}\end{lrbox}%
   \vspace{6pt}\begingroup\setlength{\fboxsep}{0pt}\colorbox{gray!20}{\usebox{\coloredbgbox}}\endgroup\vspace{12pt}}

\newcommand{\programline}[2]{%
	\emph{#1:}&#2
}

\begin{document}

\pagestyle{empty}
\hfill AEI-2013-236
\vskip 0.15\textheight
\begin{center}
{\bfseries 
	{\Huge xTras} \\ 
	\vskip 0.5cm
	{\Large a field-theory inspired \xAct~package for Mathematica}
}

\vskip 0.12\textheight

{\bfseries Teake Nutma}%

\vskip 0.5cm

\emph{\href{http://www.aei.mpg.de/}{Max-Planck-Institut f\"ur
Gravitationsphysik}}\\
\emph{(Albert Einstein Institut)}\\ 
\emph{Am M\"uhlenberg 1, } \\ 
\emph{14476 Golm, Germany}	\\ 
\vskip 0.3cm

\vskip 0.3cm
{\tt teake.nutma@aei.mpg.de}%

\end{center}

\vskip 0.08\textheight

\begin{center} {\bf Abstract } 
\end{center}
\textnohyphenation{
\begin{quotation}\noindent
We present the tensor computer algebra package \xTras, which provides functions and methods
frequently needed when doing (classical) field theory. Amongst others, it can compute contractions,
make Ans\"atze, and solve tensorial equations. It is built upon the tensor computer algebra
system \xAct, a collection of packages for Mathematica.
\end{quotation}
}

\begin{center} {\bf Program summary} 

\begin{tabu}{@{}ll}
& \\
\programline{Program title}{xTras} \\
\programline{Version}{1.2.1} \\
\programline{Programming language}{Mathematica (version 6.0 and higher)} \\
\programline{Obtainable from}{\href{http://www.xact.es/xtras/}{\texttt{www.xact.es/xtras}}} \\
\programline{License}{GPL}
\end{tabu}

\end{center}

\newpage
\pagestyle{plain}

\tableofcontents

\section{Introduction}

\xAct~\cite{MartinGarcia:2002xa} is a free collection of powerful Mathematica packages for tensor 
computer algebra. Thanks to its implementation \cite{MartinGarcia:2008,MartinGarcia:2002} 
of the Butler-Portugal algorithm \cite{Portugal:1998qi, Portugal:1999, Manssur:2002}, 
it can canonicalize tensor indices with respect to permutation symmetries extremely fast. On this
solid basis a great number of applications have been build \cite{Brizuela:2008ra,Brizuela:2006,MartinGarcia:2007ab,
MartinGarcia:2008qz,GomezLobo:2011xv,Backdahl:2013,Pitrou:2013hga} that range from tensor spherical harmonics
to perturbations around homogeneous cosmological backgrounds.

This paper describes the \xTras~package, one of these applications. \xTras~provides functions
and methods that are frequently needed when doing (classical) field theory: computing contractions,
making Ans\"atze, and solving equations, just to name a few. The package grew out of a need of the 
author for these particular functions, which were not present in any other \xAct~package.%
\footnote{
Some of the functionality of \xTras~did however already exist in another computer algebra system, 
namely Cadabra \cite{Peeters:2007wn}.
}

This paper is organized as follows. \hyperref[sec:installation]{Section \ref*{sec:installation}}
describes how to install and run
the package, \autoref{sec:xtensor} briefly reviews the basics usage of \xAct, 
\autoref{sec:examples} demonstrates of capabilities of \xTras~with a couple of examples,
and \autoref{sec:functions} contains function documentation. In addition to this paper,
a complete list of all functions and their options can be found in either the built-in
documentation of the package, or the online documentation at 
\href{http://www.xact.es/xtras/documentation/}{\texttt{www.xact.es/xtras/documentation}}.

\section{Installation}
\label{sec:installation}

\xTras~can be installed by downloading the package from its website 
\href{http://www.xact.es/xtras/}{\texttt{www.xact.es/xtras}}, 
unzipping it, and following the supplied instructions. Once installed, \xTras~can be loaded with the following command:
{\small \ttfamily
\begin{longtabu}{@{\;\;\;}r@{\;\;}X@{}}
	\normalfont{\textcolor{darkgray}{\;\;In\,:=}} & {\normalsize\bfseries <$\mspace{0.5mu}$<xAct\textasciigrave xTras\textasciigrave } \\
& ------------------------------------------------------------\\
& Package xAct\textasciigrave xPerm\textasciigrave~version 1.2.0, \{2013,1,27\}\\
& CopyRight (C) 2003-2013, Jose M. Martin-Garcia, under the General Public License.\\
& ------------------------------------------------------------\\
& Package xAct\textasciigrave xTensor\textasciigrave~version 1.0.5, \{2013,1,30\}\\
& CopyRight (C) 2002-2013, Jose M. Martin-Garcia, under the General Public License.\\
& ------------------------------------------------------------\\
& Package xAct\textasciigrave xPert\textasciigrave~version 1.0.3, \{2013,1,27\}\\
& CopyRight (C) 2005-2013, David Brizuela, Jose M. Martin-Garcia and Guillermo A. Mena Marugan, under the General Public License.\\
& ------------------------------------------------------------\\
& Package xAct\textasciigrave Invar\textasciigrave~version 2.0.4, \{2013,1,27\}\\
& CopyRight (C) 2006-2013, J. M. Martin-Garcia, D. Yllanes and R. Portugal, under the General Public License.\\
& ------------------------------------------------------------\\
& Package xAct\textasciigrave xCoba\textasciigrave~version 0.8.0, \{2013,1,30\}\\
& CopyRight (C) 2005-2013, David Yllanes and Jose M. Martin-Garcia, under the General Public License.\\
& ------------------------------------------------------------\\
& Package xAct\textasciigrave SymManipulator\textasciigrave~version 0.8.5, \{2013,4,13\}\\
& CopyRight (C) 2011-2013, Thomas B\"ackdahl, under the General Public License.\\
& ------------------------------------------------------------\\
& Package xAct\textasciigrave xTras\textasciigrave~version 1.1.1, \{2013,1,26\}\\
& CopyRight (C) 2012-2013, Teake Nutma, under the General Public License.\\
& ------------------------------------------------------------
\end{longtabu}
}
This loaded not only \xTras, but also all other \xAct~packages that
it depends on: \emph{xPerm} \cite{MartinGarcia:2008}, \emph{xTensor} \cite{MartinGarcia:2002}, 
\emph{xPert} \cite{Brizuela:2008ra}, \emph{Invar} \cite{MartinGarcia:2007ab,MartinGarcia:2008qz}, 
\emph{xCoba} \cite{Yllanes:2013}, and \emph{SymManipulator} \cite{Backdahl:2013}. Note that we have
suppressed some print messages in the Mathematica output above, and have only shown the package info.
In the rest of this paper, all print message will be suppressed.

\newsavebox\usageinfo
\savebox\usageinfo{%
\taburowcolors 2{gray!20 .. gray!20}
\tabulinesep = 8pt%
\begin{tabu}  {l}
	\hline
    MakeTraceless[\textit{expr}] returns the traceless version of \textit{expr}. \small{>\!\!>}
  \end{tabu}%
}

Once \xTras~is loaded, the built-in documentation may be opened with the command
\begin{mma}
	\mmain{xTrasHelp[]}
\end{mma}%
or alternatively by first opening Mathematica's \emph{Documentation Center} by pressing F1 and 
then searching for ``xTras''. 
Furthermore, information about \xTras~functions can be displayed, like all regular
Mathematica functions, by typing \func{?\,functionname}. For example,
\begin{mma}
	\mmain{?\;MakeTraceless} \\
	\mmanoout{\usebox{\usageinfo}}
\end{mma}%
gives a brief description of the function \funcref{MakeTraceless}. Pressing the \func{>\!>} link 
opens its help page where more detailed documentation can be found.

\section{xTensor basics}
\label{sec:xtensor}

Before we discuss \emph{xTras}, it is convenient to go over the basics of \emph{xTensor} 
\cite{MartinGarcia:2002}. 
The \emph{xTensor} package is more or less the cornerstone of \xAct, as it implements the basic 
structures of manifolds, tensors, and Riemannian geometry.

\setcounter{table}{0}
\begin{table}[H]
\begingroup\setlength{\fboxsep}{0pt}\colorbox{gray!20}{%
\begin{tabu} to \linewidth {llXl}
\hline
	&&& \\[-6pt]
	& {\ttfamily DefManifold[{\it M}, {\it d}, \{$i_1,i_2,\ldots,i_n$\}]} & 
	defines the \textit{d}-dimensional manifold \textit{M} whose tensors will have indices 
	$i_1,i_2,\ldots,i_n$. & \\[8pt]
	& {\ttfamily DefMetric[{\it sign}, {\it g}[$i_1$,$i_2$], {\it cd}]} & 
	defines a metric \textit{g} of signature \textit{sign} on the manifold of which $i_1$ and 
	$i_2$ are indices, a covariant derivative \textit{cd}, and all curvature tensor of \textit{g}. 
	& \\[8pt]
	& {\ttfamily DefTensor[{\it T}[$i_1$,\ldots,$i_m$], {\it M}, {\it sym}]} & 
	defines a tensor \textit{T} with indices $i_1$,\ldots,$i_m$ and symmetry \textit{sym} on the manifold \textit{M}. 
	& \\[8pt]
	& {\ttfamily ContractMetric[{\it expr}]} & 
	contracts all metrics in \textit{expr}. 
	& \\[4pt]
	& {\ttfamily ToCanonical[{\it expr}]} & 
	canonicalizes all tensors in \textit{expr}. 
	& \\[4pt]
	\hline
\end{tabu}%
}\endgroup
\caption{Basic commands in \emph{xTensor}.}
\end{table}

The first step in any \emph{xAct}~calculation is always to define a manifold. This can be done
with the aptly named command \func{DefManifold}:
\begin{mma}
	\mmain{DefManifold[M,~4,~IndexRange[a,m]]}
\end{mma}%
The first argument is the name of the manifold, in this case \func{M}. The second is its dimension.
This either has to be an integer or a constant symbol (which needs to be defined as such with the 
command \func{DefConstantSymbol}). The last argument of \func{DefManifold} specifies the indices
which will be used by tensors on the manifold; here \func{IndexRange[a,m]} is a convenient
short-hand for \func{\{a,b,c,d,e,f,g,h,i,j,k,l,m\}}.

After defining a manifold, it is possible to define a metric on that manifold with the command 
\func{DefMetric}:
\begin{mma}
	\mmain{DefMetric[-1, metric[-a,-b], CD, PrintAs -> "g"]}
\end{mma}%
This defined a metric \func{metric} of signature $-1$ on the manifold \func{M}, because $a$ and $b$
are indices of \func{M}. Note that we could not use \func{g} for the name of the metric, because 
\func{g} is already an index. The option \func{PrintAs} makes sure that every time the metric 
appears in any output, it gets printed as \func{g}:
\begin{mma}
	\mmain{metric[-a,~-b]}\\ 
	\mmaout{g_{ab}}
\end{mma}%
The minus signs in front of the indices indicate that they're covariant indices. 
Indices without a minus sign are contravariant:
\begin{mma}
	\mmain{metric[a,~b]}\\ 
	\mmaout{g^{ab}}
\end{mma}%
Besides defining a metric, the command \func{DefMetric} also defined curvature tensors, like
for instance the Riemann tensor,
\begin{mma}
	\mmain{RiemannCD[-a,~-b,~-c,~-d]}\\ 
	\mmaout{R_{abcd}}
\end{mma}%
and the Ricci tensor:
\begin{mma}
	\mmain{RicciCD[-a,~-b]}\\ 
	\mmaout{R_{ab}}
\end{mma}%
Their name indicates that they're associated to the covariant derivative \func{CD}, which also
has been defined:
\begin{mma}
	\mmain{CD[-a][RicciCD[-c,~-d]]}\\ 
	\mmaout{\triangledown_{a}R_{cd}}
\end{mma}%
By default, \func{DefMetric} defines a torsionless and metric compatible connection, and uses the conventions
$[\nabla_a,\nabla_b] T_c  = R_{abc}{}^d T_d$ and $R_{ab}  = R_{acb}{}^c$ for the curvature tensors.
Contractions of the Riemann tensor are automatically converted to Ricci tensors%
\footnote{
	This behavior is actually controlled by the option \func{CurvatureRelations} of \func{DefMetric} 
	(and \func{DefCovD}), which defaults to \func{True}. Torsion can be turned on by setting the option 
	\func{Torsion} to \func{True}, 
	and the relative signs for the Riemann and Ricci tensors are set via 
	the global variables \func{\$RiemannSign} and \func{\$RicciSign}.
}:
\begin{mma}
	\mmain{RiemannCD[-c,~-b,~-a,~b]}\\ 
	\mmaout{R_{ca}}
\end{mma}%
But contractions with an explicit metric are not converted:
\begin{mma}
	\mmain{metric[b,~d]~RiemannCD[-c,~-b,~-a,~-d]}\\ 
	\mmaout{g^{bd} R_{cbad}}
\end{mma}%
This is because \emph{xTensor} does not automatically contract metrics. Contracting metrics can
be done with the command \func{ContractMetric}, which does as its name suggests:
\begin{mma}
	\mmain{metric[a,~c]~RicciCD[-c,~-b]~//~ContractMetric}\\ 
	\mmaout{R^{a}{}_{b}}
\end{mma}%
And indeed, applying \func{ContractMetric} to the previous example gives the Ricci tensor:
\begin{mma}
	\mmain{metric[b,~d]~RiemannCD[-c,~-b,~-a,~-d]~//~ContractMetric}\\ 
	\mmaout{R_{ca}}
\end{mma}%
Note, however, that \emph{xTensor} also does not automatically rewrite $R_{ca}$ to $R_{ac}$,
even though the Ricci tensor is symmetric. To achieve this, we have to use the function
\func{ToCanonical}:
\begin{mma}
	\mmain{RicciCD[-c,~-a]~//~ToCanonical}\\ 
	\mmaout{R_{ac}}
\end{mma}%
Very loosely speaking, \func{ToCanonical} tries to sort indices as much as possible based on the
symmetries of the tensors in the expression (see \cite{MartinGarcia:2008} for more details). 
Needless to say, it also works on more complicated expressions:
\begin{mma}
	\mmain{RicciCD[c,~d]~RiemannCD[-d,~-b,~-a,~-c]~//~ToCanonical}\\ 
	\mmaout{- R^{cd} R_{acbd}}
\end{mma}%
At the moment, \func{ToCanonical} only simplifies so-called \emph{mono-term} symmetries, which are of the
form $T_{i_1 \cdots i_n} = \pm T_{\sigma(i_1 \cdots i_n)}$, where $\sigma \in S_n$ is a permutation of
the indices. It does not simplify so-called \emph{multi-term} symmetries, which are of the form
$T_{i_1 \cdots i_n} = \pm T_{\sigma_1(i_1 \cdots i_n)} \pm T_{\sigma_2(i_1 \cdots i_n)} + \cdots$.
One example of a multi-term symmetry is for instance the Bianchi identity $R_{[abc]d}=0$. 

After having covered the very basics of \emph{xTensor}, we are now ready to tackle more advanced examples
with the help of functions in \xTras.

\section{Examples}
\label{sec:examples}

We will now demonstrate the features of \xTras, or at least some of them, on the basis of two
examples. The functions used here are described in more detail in \autoref{sec:functions}.

\subsection{Spin 2 on a flat background}
\label{ex:spin2}

In this section we will construct a gauge invariant theory of a free spin 2 field on a flat 
background. In doing so, we will recover the linearized Einstein equations.

\begin{table}[H]
\begingroup\setlength{\fboxsep}{0pt}\colorbox{gray!20}{%
\begin{tabu} to \linewidth {llXl}
\hline
	&&& \\[-6pt]
	& {\ttfamily \funcref{AllContractions}[{\it expr}]} & 
	computes all possible contractions of \textit{expr}. & \\
	&&& \\[-6pt]
	& {\ttfamily \funcref{MakeAnsatz}[\{$e_1, e_2, \ldots$\}]} & 
	makes an Ansatz out of the list entries $e_1, e_2, \ldots$. 
	& \\
	&&& \\[-6pt]
	& {\ttfamily \funcref{CollectTensors}[{\it expr}]} &  
	groups all tensorial terms in \textit{expr} together. 
	& \\
	&&& \\[-6pt]
	& {\ttfamily \funcref{SolveConstants}[{\it expr}]} & 
	attempts to solve the system \textit{expr} of tensorial equations & \\
	& &  for all constant symbols appearing in \textit{expr}.
	& \\[4pt]
	\hline
\end{tabu}%
}\endgroup
\caption{
	\xTras~functions used in this example. 
	They are described in more detail in \autoref{sec:functions}.
}
\end{table}

After loading the package, we have to define a manifold and a flat metric. This can be done
as follows:
\begin{mma}
	\mmain{DefManifold[M,~dim,~IndexRange[a,m]];} \\
	\mmain{DefMetric[}\\[-5pt]
	\mmanoin{\tab -1,~metric[-a,-b],~PD,~PrintAs -> "$\eta$",} \\[-5pt]
	\mmanoin{\tab FlatMetric -> True,~SymbolOfCovD -> \{",","$\partial$"\}} \\[-5pt]
	\mmanoin{]}
\end{mma}%
This did not define a new covariant derivative, but instead set the pre-existing 
partial derivative \func{PD} to be metric compatible with \func{metric}.
Furthermore, we need to tell the function \func{SymmetryOf} 
that the metric is constant:
\begin{mma}
	\mmain{SetOptions[SymmetryOf, ConstantMetric -> True];} \\
\end{mma}%
Besides defining a manifold and a metric, we also need to define a symmetric spin two field and a gauge vector:
\begin{mma}
	\mmain{DefTensor[H[-a, -b], M, Symmetric[\{-a, -b\}], PrintAs -> "h"];}\\
	\mmain{DefTensor[xi[a], M, PrintAs -> "$\xi$"] }
\end{mma}%
We are now ready to begin the actual computation. We will construct all possible terms for the action,
and make an Ansatz out of them. Because we are not interested in total derivatives, 
it suffices to consider terms of the form $h \cdot \partial \cdot \partial \cdot h$. First, find all
of these terms:
\begin{mma}
	\mmain{Sterms~=~AllContractions[~H[a,~b]~PD[c]@PD[d]@H[e,~f]~]}\\ 
	\mmaout{\texttt{\{}h^{ab} \partial_{b}\partial_{a}h^{c}{}_{c}\texttt{, }h^{ab} \partial_{c}\partial_{b}h_{a}{}^{c}\texttt{, }h^{a}{}_{a} \partial_{c}\partial_{b}h^{bc}\texttt{, }h^{ab} \partial_{c}\partial^{c}h_{ab}\texttt{, }h^{a}{}_{a} \partial_{c}\partial^{c}h^{b}{}_{b}\texttt{\}}}
\end{mma}%
Now construct the action:
\begin{mma}
	\mmain{S~=~MakeAnsatz[Sterms]}\\ 
	\mmaout{C_1 h^{ab} \partial_{b}\partial_{a}h^{c}{}_{c} + C_2 h^{ab} \partial_{c}\partial_{b}h_{a}{}^{c} + C_3 h^{a}{}_{a} \partial_{c}\partial_{b}h^{bc} + C_4 h^{ab} \partial_{c}\partial^{c}h_{ab} + C_5 h^{a}{}_{a} \partial_{c}\partial^{c}h^{b}{}_{b}}
\end{mma}%
The equations of motion are then:
\begin{mma}
	\mmain{eom~=~VarD[H[-a,~-b],~PD][S]~//~CollectTensors}\\ 
	\mmaout{(C_1 + C_3^{\text{}}) \partial^{b}\partial^{a}h^{c}{}_{c} + C_2 \partial_{c}\partial^{a}h^{bc} + C_2 \partial_{c}\partial^{b}h^{ac} + 2 C_4 \partial_{c}\partial^{c}h^{ab} + (C_1 + C_3^{\text{}}) \eta^{ab} \partial_{d}\partial_{c}h^{cd}}\\ 
	\mmanoout{ + 2 C_5 \eta^{ab} \partial_{d}\partial^{d}h^{c}{}_{c}}
\end{mma}%
We want to make the action and the equations of motion gauge invariant under the following gauge transformation 
of the spin two field:
\begin{mma}
	\mmain{$\delta$H~=~2~Symmetrize[ PD[-a]@xi[-b] ]}\\ 
	\mmaout{\partial_{a}\xi_{b} + \partial_{b}\xi_{a}}
\end{mma}%
To that end, we compute the gauge variation of the action $\delta S$ to be
\begin{mma}
	\mmain{$\delta$S~=~$\delta$H~eom~//~CollectTensors}\\ 
	\mmaout{2 (C_1 + C_3^{\text{}}) \partial_{b}\partial_{a}h^{c}{}_{c} \partial^{b}\xi^{a} + 2 C_2 \partial^{b}\xi^{a} \partial_{c}\partial_{a}h_{b}{}^{c} + 2 C_2 \partial^{b}\xi^{a} \partial_{c}\partial_{b}h_{a}{}^{c} + }\\ 
	\mmanoout{2 (C_1 + C_3^{\text{}}) \partial_{a}\xi^{a} \partial_{c}\partial_{b}h^{bc} + 4 C_4 \partial^{b}\xi^{a} \partial_{c}\partial^{c}h_{ab} + 4 C_5 \partial_{a}\xi^{a} \partial_{c}\partial^{c}h^{b}{}_{b}}
\end{mma}%
Up to total derivatives, this should be zero. We can eliminate total derivatives by removing all
derivatives from the gauge parameter with the help of \func{VarD}:
\begin{mma}
	\mmain{$\delta$S~=~xi[a]~VarD[xi[a],~PD][$\delta$S]~//~CollectTensors}\\ 
	\mmaout{-2 (C_1 + C_2 + C_3) \xi^{a} \partial_{c}\partial_{b}\partial_{a}h^{bc} - 2 (C_1 + C_3 + 2 C_5) \xi^{a} \partial_{c}\partial^{c}\partial_{a}h^{b}{}_{b}}\\
	\mmanoout{- 2 (C_2 + 2 C_4) \xi^{a} \partial_{c}\partial^{c}\partial_{b}h_{a}{}^{b}}
\end{mma}%
Finally, we demand the above to be zero by solving for the unknown constants:
\begin{mma}
	\mmain{sols~=~SolveConstants[$\delta$S~==~0]}\\ 
	\mmaout{\texttt{\{}\texttt{\{}C_3^{\text{}} \rightarrow - C_1 -  C_2^{\text{}}\texttt{, }C_4^{\text{}} \rightarrow - \tfrac{1}{2} C_2^{\text{}}\texttt{, }C_5^{\text{}} \rightarrow \tfrac{1}{2} C_2^{\text{}}\texttt{\}}\texttt{\}}}
\end{mma}%
Plugging this solution into the action, we find
\begin{mma}
	\mmain{S~/.~First[sols]}\\ 
	\mmaout{C_1 h^{ab} \partial_{b}\partial_{a}h^{c}{}_{c} + C_2 h^{ab} \partial_{c}\partial_{b}h_{a}{}^{c} + (- C_1 -  C_2) h^{a}{}_{a} \partial_{c}\partial_{b}h^{bc} -  \tfrac{1}{2} C_2 h^{ab} \partial_{c}\partial^{c}h_{ab}}\\ 
	\mmanoout{ + \tfrac{1}{2} C_2 h^{a}{}_{a} \partial_{c}\partial^{c}h^{b}{}_{b}}
\end{mma}%
The coefficient $C_2$ parameterizes an overall normalization, and the coefficient $C_1$ a total
derivative. Indeed, $C_1$ does not appear in the final equations of motion:
\begin{mma}
	\mmain{eom~/.~First[sols]~/.~C2~->~1}\\ 
	\mmaout{- \partial^{b}\partial^{a}h^{c}{}_{c} + \partial_{c}\partial^{a}h^{bc} + \partial_{c}\partial^{b}h^{ac} -  \partial_{c}\partial^{c}h^{ab} -  \eta^{ab} \partial_{d}\partial_{c}h^{cd} + \eta^{ab} \partial_{d}\partial^{d}h^{c}{}_{c}}
\end{mma}%
These are precisely the linearized Einstein equations.

\subsection{Gauss-Bonnet term}

In this section we will show that the Euler density in four dimensions, also known as the
Gauss-Bonnet term, is topological. That is, we will show that its equations of motion vanish
identically.

\begin{table}[H]
\begingroup\setlength{\fboxsep}{0pt}\colorbox{gray!20}{%
\begin{tabu} to \linewidth {llXl}
\hline
	&&& \\[-6pt]
	& {\ttfamily \funcref{EulerDensity}[{\it cd}]} & 
	gives the Euler density associated to the covariant & \\
	& & derivative \textit{cd}. & \\[1pt]
	& {\ttfamily \funcref{VarL}[{\it g}[-{\it a},-{\it b}]][{\it L}]} & 
	computes $\tfrac{1}{\sqrt{\lvert g \rvert}}\tfrac{\delta \sqrt{\lvert g \rvert} L}{\delta g_{ab}}$.
	& \\
	&&& \\[-7pt]
	& {\ttfamily \funcref{FullSimplification}[][{\it expr}]} &  
	tries to simplify \textit{expr} as much as possible, taking & \\
	& &  Bianchi identities into account and sorting covariant & \\
	& & derivatives.
	& \\[4pt]
	& {\ttfamily \funcref{ConstructDDIs}[{\it expr}]} &  
	constructs all scalar dimensional dependent & \\
	& & identities that can be build out of \textit{expr}.
	& \\[4pt]
	& {\ttfamily \funcref{SolveTensors}[{\it expr}]} & 
	attempts to solve the system \textit{expr} of tensorial & \\
	& & equations for all tensors in \textit{expr}.
	& \\[4pt]
	\hline
\end{tabu}%
}\endgroup
\caption{
	New \xTras~functions used in this example. 
	They are described in more detail in \autoref{sec:functions}.
}
\end{table}

\noindent We will begin from scratch, and define a manifold and metric:
\begin{mma}
	\mmain{DefManifold[M,~4,~IndexRange[a,m]];} \\
	\mmain{DefMetric[-1,~metric[-a,-b],~CD,~PrintAs->"g"];}
\end{mma}%
Next, we determine the Gauss-Bonnet term via the function \funcref{EulerDensity}:
\begin{mma}
	\mmain{GBterm~=~NoScalar~@~EulerDensity[CD]}\\ 
	\mmaout{4 R_{ab} R^{ab} -  R^2 -  R_{abcd} R^{abcd}}
\end{mma}%
The \func{NoScalar} call removed any \func{Scalar} heads in the expression (see also \autoref{EulerDensity}). 
Note that \funcref{EulerDensity}
omits the overall factor $\sqrt{-g}$, so technically speaking \func{GBterm} is not a density.
The equations of motion of the Gauss-Bonnet term can be determined with the function \funcref{VarL},
and simplified with \funcref{FullSimplification}:
\begin{mma}
	\mmain{eom~=~FullSimplification[]~@~VarL[metric[-a,~-b]]~@~GBterm}\\ 
	\mmaout{-4 R^{ac} R^{b}{}_{c} + 2 g^{ab} R_{cd} R^{cd} + 2 R^{ab} R -  \tfrac{1}{2} g^{ab} R^2 - 4 R^{cd} R^{a}{}_{c}{}^{b}{}_{d} + 2 R^{acde} R^{b}{}_{cde} }\\
	\mmanoout{-  \tfrac{1}{2} g^{ab} R_{cdef} R^{cdef}}
\end{mma}%
Because the Gauss-Bonnet term is topological, the above should identically be zero. There are no
further simplifications coming from Bianchi identities that we can use: 
\hyperref[FullSimplification]{\func{FullSimplifi-}} \hyperref[FullSimplification]{\func{cation}} took
care of most of them, and if there were some remaining we still could not use them to get rid of the
Ricci tensors.

So the above equations of motion can only be zero due to dimensionally dependent identities.
We can obtain the relevant identities with a call to \funcref{ConstructDDIs}:
\begin{mma}
	\mmain{ddis~=~ConstructDDIs[}\\[-5pt]
	\mmanoin{\tab RiemannCD[a,~b,~c,~d]~RiemannCD[e,~f,~g,~h],}\\[-5pt]
	\mmanoin{\tab \{a,~b\}}\\[-5pt]
	\mmanoin{]}\\ 
	\mmaout{\texttt{\{}R^{ac} R^{b}{}_{c} -  \tfrac{1}{2} g^{ab} R_{cd} R^{cd} -  \tfrac{1}{2} R^{ab} R + \tfrac{1}{8} g^{ab} R^2 + R^{cd} R^{a}{}_{c}{}^{b}{}_{d} -  R^{acde} R^{b}{}_{cde} + R^{acde} R^{b}{}_{dce}}\\
	\mmanoout{\phantom{\texttt{\{}} + \tfrac{1}{4} g^{ab} R_{cdef} R^{cdef} -  \tfrac{1}{4} g^{ab} R_{cedf} R^{cdef}\texttt{, }}\\ 
	\mmanoout{\phantom{\texttt{\{}}R^{acde} R^{b}{}_{cde} - 2 R^{acde} R^{b}{}_{dce} -  \tfrac{1}{4} g^{ab} R_{cdef} R^{cdef} + \tfrac{1}{2} g^{ab} R_{cedf} R^{cdef}\texttt{,}} \\
	\mmanoout{\phantom{\texttt{\{}}R^{ac} R^{b}{}_{c} -  \tfrac{1}{2} g^{ab} R_{cd} R^{cd} -  \tfrac{1}{2} R^{ab} R + \tfrac{1}{8} g^{ab} R^2 + R^{cd} R^{a}{}_{c}{}^{b}{}_{d} -  \tfrac{1}{2} R^{acde} R^{b}{}_{cde} + \tfrac{1}{8} g^{ab} R_{cdef} R^{cdef}\texttt{, }}\\
	\mmanoout{\phantom{\texttt{\{}}R^{ac} R^{b}{}_{c} -  \tfrac{1}{2} g^{ab} R_{cd} R^{cd} -  \tfrac{1}{2} R^{ab} R + \tfrac{1}{8} g^{ab} R^2 + R^{cd} R^{a}{}_{c}{}^{b}{}_{d} -  R^{acde} R^{b}{}_{dce} + \tfrac{1}{4} g^{ab} R_{cedf} R^{cdef}\texttt{\}}}
\end{mma}%
This constructed all dimensionally dependent identities that have two Riemann tensors
(or contractions thereof) and free indices $a$ and $b$. All of these four expression are zero.
Even though there are four identities,
only two of them are independent (not taking Bianchi identities into account). 
This can be verified with \funcref{SolveTensors}:
\begin{mma}
	\mmain{ddisols~=~SolveTensors[}\\[-5pt]
	\mmanoin{\tab ddis~==~0,}\\[-5pt]
	\mmanoin{\tab UseSymmetries~->~False,~MetricOn~->~None}\\[-5pt]
	\mmanoin{]}
\end{mma}
\begin{mma*}
	\mmaout{\texttt{\{}\texttt{\{}\texttt{HoldPattern[}R{}^{\uunderline{acde}} R{}^{\uunderline{b}}{}_{\uunderline{cde}}\texttt{]}~\ruledelayed \texttt{Module[}\texttt{\{}f\texttt{, }h\texttt{, }i\texttt{, }g\texttt{, }j\texttt{, }k\texttt{, }l\texttt{, }m\texttt{\}, }}\\
	\mmanoout{\tab\tab 2 R^{af} R^{b}{}_{f} -  g^{ab} R_{hi} R^{hi} -  R^{ab} R + \tfrac{1}{4} g^{ab} R^2 + 2 R^{fg} R^{a}{}_{f}{}^{b}{}_{g} + \tfrac{1}{4} g^{ab} R_{jklm} R^{jklm}\texttt{]}\texttt{, }}\\
	\mmanoout{\phantom{\texttt{\{}\texttt{\{}}\texttt{HoldPattern[}R{}^{\uunderline{acde}} R{}^{\uunderline{b}}{}_{\uunderline{dce}}\texttt{]}~\ruledelayed \texttt{Module[}\texttt{\{}f\texttt{, }h\texttt{, }i\texttt{, }g\texttt{, }j\texttt{, }k\texttt{, }l\texttt{, }m\texttt{\}, }}\\
	\mmanoout{\tab\tab R^{af} R^{b}{}_{f} -  \tfrac{1}{2} g^{ab} R_{hi} R^{hi} -  \tfrac{1}{2} R^{ab} R + \tfrac{1}{8} g^{ab} R^2 + R^{fg} R^{a}{}_{f}{}^{b}{}_{g} + \tfrac{1}{4} g^{ab} R_{jlkm} R^{jklm}\texttt{]}\texttt{\}}\texttt{\}}}
\end{mma*}%
The two options are needed to prevent \func{SolveTensors} from making rules for every index combination
on the left-hand-side related by symmetries (\func{UseSymmetries}) and by raising and lowering of 
the indices (\func{MetricOn}). Because \funcref{SolveTensors} returns a solution for two tensor
structures in terms of others, only two of the four found DDIs are independent.%
\footnote{
	Again, this is true up to Bianchi identities. If we take those into account, there is only one
	truly independent DDI because $R_{acde} R^{bdce} = \frac{1}{2} R_{acde} R^{bcde}$. This 
	identity is derived in \autoref{RiemannYoungProject}.
}

The above output consists of rules that we can use to enforce the identities on the equations of motion:
\begin{mma}
	\mmain{eom~/.~ddisols~//~ToCanonical}\\ 
	\mmaout{\texttt{\{}0\texttt{\}}}
\end{mma}%
So, indeed, the equations of motion are zero. 

Alternatively, we could have made an Ansatz with arbitrary coefficients from the identities, 
\begin{mma}
	\mmain{ddiAnsatz~=~CollectTensors~@~MakeAnsatz[ddis]}\\ 
	\mmaout{(C_1 + C_2 + C_3) R^{ac} R^{b}{}_{c} + \tfrac{1}{2} (- C_1 -  C_2 -  C_3) g^{ab} R_{cd} R^{cd} + \tfrac{1}{2} (- C_1 -  C_2 -  C_3) R^{ab} R }\\
	\mmanoout{+ \tfrac{1}{8} (C_1 + C_2 + C_3) g^{ab} R^2 + (C_1 + C_2 + C_3) R^{cd} R^{a}{}_{c}{}^{b}{}_{d} + (- \tfrac{1}{2} C_1 -  C_2 + C_4) R^{acde} R^{b}{}_{cde}}\\ 
	\mmanoout{  + (C_2 -  C_3 - 2 C_4) R^{acde} R^{b}{}_{dce} + \tfrac{1}{8} (C_1 + 2 C_2 - 2 C_4) g^{ab} R_{cdef} R^{cdef}}\\
	\mmanoout{ + \tfrac{1}{4} (- C_2 + C_3 + 2 C_4) g^{ab} R_{cedf} R^{cdef}}
\end{mma}%
and tried to make this equal to the equations of motion by solving for the coefficients:
\begin{mma}
	\mmain{SolveConstants[eom~==~ddiAnsatz]}\\ 
	\mmaout{\texttt{\{}\texttt{\{}C_3^{\text{}} \rightarrow -4 -  C_1 -  C_2^{\text{}}\texttt{, }C_4^{\text{}} \rightarrow 2 + \tfrac{1}{2} C_1 + C_2^{\text{}}\texttt{\}}\texttt{\}}}
\end{mma}%
So again, the equations of motion are equal to particular linear combination of the dimensionally
dependent identities, and hence they are zero.

\section{xTras functions}
\label{sec:functions}

This section documents the most important functions in \xTras. The list of functions below is not
exhaustive, nor are the functions described in full detail (for example, most options are 
not described here).
For a complete list of functions and all their options, please refer to the built-in
documentation or the online documentation at 
\href{http://www.xact.es/xtras/documentation/}{\texttt{www.xact.es/xtras/documentation}}.

Throughout this section, we assume we have a manifold \func{M}, a metric \func{metric}, a covariant 
derivative \func{CD} and associate curvature tensors (\func{RiemannCD}, \func{RicciCD}, etc).
These can be defined with the commands
\begin{mma}
	\mmain{DefManifold[M,~dim,~IndexRange[a,m]];} \\
	\mmain{DefMetric[-1,~metric[-a,-b],~CD,~PrintAs->"g"];}
\end{mma}%
where \func{dim} is a predefined constant symbol.

\subsection{Combinatorics}

In this section we discuss some of the \xTras~functions that are of a combinatorial nature.

\subsubsection{AllContractions}
\label{AllContractions}

\begin{usagetable}
	\usageline
		{AllContractions[{\it expr}]}
		{returns a sorted list of all possible full contractions of \textit{expr} over its free indices.}
	\usageline
		{AllContractions[{\it expr}, {\it frees}]}
		{returns all possible contractions of \textit{expr} that have \textit{frees} as free indices.}
	\usagelinelong
		{AllContractions[{\it expr}, {\it frees}, {\it sym}]}
		{returns all possible contractions of \textit{expr} with the symmetry \textit{sym} imposed on the free indices \textit{frees}.}
\end{usagetable}

\paragraph{Details}

A recurring problem in field theory is to make the most general Ansatz that contains a specific set
of fields and derivatives. For constructing for example the most general gauge-invariant action for
a particular set of fields (like we did for the free spin-2 field on a flat background in \autoref{ex:spin2}), 
one would need to know all possible vertices and all
possible gauge transformations. While this problem is still tractable at lowest orders, it becomes
complicated very fast at higher orders. In fact, the naive number of possible contractions of a
tensorial expression that has $n$ free indices is $(n-1)!!$, which is the number of independent
products of $\frac{n}{2}$ metrics. The problem of finding all contractions when $n$ is large is, if not error-prone,
tedious at the very least. That's where  the command \func{AllContractions} comes in.

The problem of finding all possible contractions of the input expression is equivalent to enumerating
all double coset representatives of $K \backslash S_n / H$, where $n$ is the number of indices of the input
expression, $K$ its symmetry group, and $H$ the symmetry group of $\tfrac n 2$ metrics. However,
double coset enumeration is known the be an NP-hard problem in general \cite{Luks:1993}, and to 
the author's knowledge no satisfactory algorithm has been found to date.

So instead of doing a proper double coset enumeration, \func{AllContractions} uses a brute-force-method to 
find all contractions. The algorithm it uses is as follows:
\begin{enumerate}
	\item Take all single contractions of the input expression.
	\item Canonicalize the single contractions, and throw away duplicates.
	\item Take all second contractions of the canonicalized single contractions.
	\item Canonicalize the second contractions, and throw away duplicates.
	\item \ldots
\end{enumerate}
\ldots and so on and so forth until all indices are contracted. This algorithm is reasonably fast
if the input expression has a large degree of symmetry, but in general it is exponential in the 
number of indices to be contracted.

\paragraph{Examples}

In its most basic form, \func{AllContractions} takes a single argument and computes all of its
possible independent full contractions.  Take for instance the Riemann tensor:
\begin{mma}
	\mmain{AllContractions[RiemannCD[-a,~-b,~-c,~-d]]}\\ 
	\mmaout{\texttt{\{}R\texttt{\}}}
\end{mma}%
As we could have expected, its only possible full contraction is the Ricci scalar. If we take two
Riemann tensors, things get a bit more interesting:
\begin{mma}
	\mmain{AllContractions[}\\[-5pt]
	\mmanoin{\tab RiemannCD[-a,~-b,~-c,~-d]~RiemannCD[-e,~-f,~-g,~-h]}\\[-5pt]
	\mmanoin{]}\\
	\mmaout{\texttt{\{}R_{ab} R^{ab}\texttt{, }R^2\texttt{, }R_{abcd} R^{abcd}\texttt{, }R_{acbd} R^{abcd}\texttt{\}}}
\end{mma}%
The last two contractions are actually not independent, but are related via the Bianchi identity.
The Bianchi identity is a multi-term symmetry, and \func{AllContractions}
does not take these symmetries into account. Hence \func{AllContractions} does not necessarily 
return an irreducible basis of contractions, but it does always return a complete basis.

It is also possible to ask for contractions of expressions with derivatives:
\begin{mma}
	\mmain{AllContractions[ RicciCD[a,~b]~CD[c]@CD[d]@RicciCD[e,~f] ]}\\ 
	\mmaout{\texttt{\{}R \triangledown_{a}\triangledown^{a}R\texttt{, }R \triangledown_{b}\triangledown_{a}R^{ab}\texttt{, }R^{ab} \triangledown_{b}\triangledown_{a}R\texttt{, }R^{ab} \triangledown_{b}\triangledown_{c}R_{a}{}^{c}\texttt{, }R^{ab} \triangledown_{c}\triangledown_{b}R_{a}{}^{c}\texttt{, }R^{ab} \triangledown_{c}\triangledown^{c}R_{ab}\texttt{\}}}
\end{mma}%
Note that besides not taking Bianchi identities into account, \func{AllContractions} also does not 
automatically sort covariant derivatives.

\func{AllContractions} takes an optional second argument, which specifies what
free indices the final contractions should have. This effectively adds an auxiliary tensor in the
first argument with the specified indices, and varies the contractions afterwards with respect to
this auxiliary tensor. Here's an example with two free indices:
\begin{mma}
	\mmain{AllContractions[}\\[-5pt]
	\mmanoin{\tab RiemannCD[-a,~-b,~-c,~-d]~RiemannCD[-e,~-f,~-g,~-h],}\\[-5pt]
	\mmanoin{\tab \{-a,~-b\}}\\[-5pt]
	\mmanoin{]}\\ 
	\mmaout{\texttt{\{}R_{a}{}^{c} R_{bc}\texttt{, }g_{ab} R_{cd} R^{cd}\texttt{, }R_{ab} R\texttt{, }g_{ab} R^2\texttt{, }R^{cd} R_{acbd}\texttt{, }R_{a}{}^{cde} R_{bcde}\texttt{, }R_{a}{}^{cde} R_{bdce}\texttt{, }}\\
	\mmanoout{\phantom{\texttt{\{}}g_{ab} R_{cdef} R^{cdef}\texttt{, }g_{ab} R_{cedf} R^{cdef}\texttt{\}}}
\end{mma}%
We can also specify an optional third argument to \func{AllContractions}. This third argument
specifies the symmetry of the indices in the second argument. For instance, we can try to
see if there are any antisymmetric contractions in the above example:
\begin{mma}
	\mmain{AllContractions[}\\[-5pt]
	\mmanoin{\tab RiemannCD[-a,-b,-c,-d]~RiemannCD[-e,-f,-g,-h], }\\[-5pt]
	\mmanoin{\tab \{-a, -b\}, }\\[-5pt]
	\mmanoin{\tab Antisymmetric[\{-a,-b\}] }\\[-5pt]
	\mmanoin{]}\\
	\mmaout{\texttt{\{\}}}
\end{mma}%
As is obvious from the previous example, there are none.

\subsubsection{MakeTraceless}
\label{MakeTraceless}

\begin{usagetable}
	\usageline
		{MakeTraceless[{\it expr}]}
		{returns the traceless version of \textit{expr}.}
\end{usagetable}

\noindent Any tensor can be projected onto its irreducible traceless components. The way to do this by
hand is to write down all possible traces of the tensor, make an Ansatz for a linear combination of them,
and then demand that single traces of this Ansatz are zero. Needless to say, for tensors of 
large rank this task is perfectly suited for the computer.

The function \func{MakeTraceless} does exactly this: it takes its argument and makes it traceless.
For the Ricci tensor, it gives the traceless Ricci tensor:
\begin{mma}
	\mmain{MakeTraceless[RicciCD[-a,~-b]]}\\ 
	\mmaout{R_{ab} -  \dfrac{g_{ab} R}{d}}
\end{mma}%
And if we enter the Riemann tensor, it returns the Weyl tensor:
\begin{mma}
	\mmain{MakeTraceless[RiemannCD[-a,~-b,~-c,~-d]]}\\ 
	\mmaout{R_{abcd} + \dfrac{2}{(d-2)(d-1)} R \, \underset{1234}{Sym}(g_{ac} g_{bd})-  \dfrac{4 }{-2 + d}\underset{1234}{Sym}(g_{bd} R_{ac})}
\end{mma}%
\func{MakeTraceless} uses the power of the \emph{SymManipulator} package \cite{Backdahl:2013} to 
implicitly impose symmetry of the Riemann tensor without expanding
all required terms. This is what the \func{Sym} objects in the above output do. We can remove them
and expand all terms with the command \func{ExpandSym}, thereby recovering 
the usual expression for the Weyl tensor:
\begin{mma}
	\mmain{MakeTraceless[RiemannCD[-a,-b,-c,-d]]~//~ExpandSym~//~ToCanonical}\\ 
	\mmaout{- \dfrac{g_{bd} R_{ac}}{-2 + d} + \dfrac{g_{bc} R_{ad}}{-2 + d} + \dfrac{g_{ad} R_{bc}}{-2 + d} -  \dfrac{g_{ac} R_{bd}}{-2 + d} -  \dfrac{g_{ad} g_{bc} R}{2 - 3 d + d^2} + \dfrac{g_{ac} g_{bd} R}{2 - 3 d + d^2} + R_{abcd}}
\end{mma}%
We can convert this to the actual Weyl tensor with the \emph{xTensor} command \func{RiemannToWeyl}:
\begin{mma}
	\mmain{RiemannToWeyl[\%]~//~ToCanonical~//~Simplify}\\ 
	\mmaout{W[\triangledown ]_{abcd}}
\end{mma}%
\func{MakeTraceless} works on any expression without dummy indices. For example, here is the traceless
version of a generic rank-3 tensor:
\begin{mma}
	\mmain{DefTensor[T[a,b,c], M];}\\ 
	\mmain{MakeTraceless[T[a,~b,~c]]}\\ 
	\mmaout{T^{abc} -  \dfrac{(1 + d) g^{bc} T^{ad}{}_{d}}{-2 + d + d^2} + \dfrac{g^{ac} T^{bd}{}_{d}}{-2 + d + d^2} + \dfrac{g^{ab} T^{cd}{}_{d}}{-2 + d + d^2} + \dfrac{g^{bc} T^{da}{}_{d}}{-2 + d + d^2} }\\ 
	\mmanoout{-  \dfrac{(1 + d) g^{ac} T^{db}{}_{d}}{-2 + d + d^2} + \dfrac{g^{ab} T^{dc}{}_{d}}{-2 + d + d^2} + \dfrac{g^{bc} T^{d}{}_{d}{}^{a}}{-2 + d + d^2} + \dfrac{g^{ac} T^{d}{}_{d}{}^{b}}{-2 + d + d^2} -  \dfrac{(1 + d) g^{ab} T^{d}{}_{d}{}^{c}}{-2 + d + d^2}}
\end{mma}%

\subsubsection{ConstructDDIs}
\label{ConstructDDIs}

\begin{usagetable}
	\usageline
		{ConstructDDIs[{\it expr}]}
		{constructs all scalar dimensional dependent identities that can be build out of \textit{expr}.}
	\usagelinelong
		{ConstructDDIs[{\it expr}, {\it frees}]}
		{constructs all dimensional dependent identities that can be build out of \textit{expr} and that have free indices \textit{frees}.}
	\usagelinelong
		{ConstructDDIs[{\it expr}, {\it frees}, {\it sym}]}
		{constructs all dimensional dependent identities that can be build out of \textit{expr} and that have the symmetry \textit{sym} imposed on their free indices \textit{frees}.}
\end{usagetable}

\paragraph{Details}

Dimensional dependent identities (DDIs) are identities that only hold in specific dimensions.
Typically, they can be derived from over--antisymmetrizations: that is, antisymmetrization over more
indices than the number of dimensions. In $d$ dimensions, one such identity is for example the 
generalized Kronecker delta with $2(d+1)$ indices:
\begin{equation}
\label{eq:fund_ddi}
	\delta_{a_1 \cdots a_{d+1}}^{\mspace{1mu}b_{\mspace{1mu}1}\cdots \mspace{1mu}b_{d+1}} = (d+1)! \, \delta_{[a_1}^{b_1} \delta_{\phantom{\negmedspace[}a_2}^{b_2} \cdots \delta_{\phantom{\negmedspace[}a_d}^{b_d} \delta_{a_{d+1}]}^{b_{d+1}} = 0.
\end{equation}
By contracting this identity with other tensors, it is possible to construct derived identities. 
For instance, in three dimensions we can contract it with a traceless $\{2,2\}$ tensor, such as the
Weyl tensor, and find
\begin{equation}
	  \delta_{[a}^e \delta_b^f \delta_c^g \delta_{d]}^h W\indices{_{gh}^{ij}} =
	 \delta_{[a}^e \delta_b^f W\indices{_{cd]}^{ij}} = 0.
\end{equation}
Contracting over $d$ and $j$ gives
\begin{equation}
	\delta_{[a}^{[e} W\indices{_{bc]}^{f]i}} = 0,
\end{equation}
and a further contraction over $i$ and $c$ gives the well-known fact that they Weyl tensor 
identically vanishes in three dimensions:
\begin{equation}
	W\indices{_{ab}^{ef}} = 0 .
\end{equation}

All DDIs that stem from over--antisymmetrizations
can in fact be derived from the `basic' identity \eqref{eq:fund_ddi} 
because it is always possible to pull out deltas on the over-antisymmetrized indices.  
Over--antisymmetrization over more than $d+1$ indices will give not give independent DDIs, because 
they can be written as linear combinations of antisymmetrizations over $d+1$ indices.

A systematic way of enumerating all DDIs is to consider all possible contractions of the fundamental
identity \eqref{eq:fund_ddi} with the relevant tensors. This is exactly what \func{ConstructDDIs}
does: it computes via \funcref{AllContractions} all contractions between the input expression and the basic identity 
\eqref{eq:fund_ddi} in the relevant dimension. 
In performing these contractions, two observations make life computationally easier: the 
independent index configurations of the basic identity are given by its standard Young tableaux,
and the basic identity is completely traceless.

The latter is important for the following reason.  While we can still write down
meaningful derived identities with the uncontracted basic identity, this is not possible with
any of its contractions -- attempting to do so results in the  trivial statement $0=0$. The
difference between the vanishing of the uncontracted and the contracted basic identity is
that the former is identically zero only when explicitly writing the indices out as 
$a,b,\ldots \in \{0,\ldots,d-1\}$,
whereas the latter is identically zero without doing so.

To see why the basic identity is traceless, consider for example the basic identity in one 
dimension:
\begin{equation}
	g_{a[b} g_{c]d} = 0.
\end{equation}
Writing out the antisymmetrization and contracting a pair of indices gives
\begin{equation}
	\tfrac{1}{2} g^{ab} \left( g_{ab} g_{cd} - g_{ac} g_{bd} \right)  
	= \frac{d-1}{2} g_{cd} = 0 , \phantom{\qquad\qquad\quad}
\end{equation}
where $d$ is the number of dimensions.
Doing the same exercise for the basic identity in two dimensions gives
\begin{equation}
	g^{ab} 
	g\indices{_{[\underline{a}b}} \, g\indices{_{\underline{c}d}} \, g\indices{_{\underline{e}]f}}
	=
	\frac{d-2}{3} g\indices{_{[\underline{c}d}} \, g\indices{_{\underline{e}]f}} 
	= 0 ,
\end{equation}
while three dimensions gives
\begin{equation}
	g^{ab} 
	g\indices{_{[\underline{a}b}} \, g\indices{_{\underline{c}d}} \, 
	g\indices{_{\underline{e}f}} \, g\indices{_{\underline{g}]h}}
	=
	\frac{d-3}{4}
	g\indices{_{[\underline{c}d}} \, g\indices{_{\underline{e}f}} \, g\indices{_{\underline{g}]h}} 
	= 0.
\end{equation}
The same holds true for other contractions. Thus the fact that the basic identity is traceless
is a dimensionally dependent statement.

The tracelessness of the basic identity allows us to only consider contractions of the form
\begin{equation}
	\delta^{a_1 \cdots a_{2(d+1)}} \left<x\right>_{a_1 \cdots a_{2(d+1)}},
\end{equation}
where $\delta^{a_1 \cdots a_{2(d+1)}}$ is the basic identity \eqref{eq:fund_ddi}, and by 
$\left<x\right>_{a_1 \cdots a_{2(d+1)}}$ we mean all contractions
of $x$ with $2(d+1)$ free indices. Taking all possible combinations of 
these contractions with the standard Young tableaux of the basic identity 
then yields all (scalar) DDIs.

\paragraph{Examples}

In two dimensions, the Einstein tensor vanishes. This can be reproduced by asking for all DDIs
with two free indices constructed out of the Riemann tensor:
\begin{mma}
	\mmain{dim = 2;}\\ 
	\mmain{ConstructDDIs[RiemannCD[a,~b,~c,~d],~\{a,~b\}]}\\ 
	\mmaout{\texttt{\{}R^{ab} -  \tfrac{1}{2} g^{ab} R\texttt{\}}}
\end{mma}%
Note that \func{ConstructDDIs} returns a list of expressions that are zero, and not equations.

In three dimensions, the Weyl tensor is zero. This time, we need four free indices that have
the symmetry of the Riemann tensor:
\begin{mma}
	\mmain{dim = 3;}\\ 
	\mmain{ConstructDDIs[}\\[-5pt]
	\mmanoin{\tab RiemannCD[a,~b,~c,~d],}\\[-5pt]
	\mmanoin{\tab \{a,~b,~c,~d\},}\\[-5pt]
	\mmanoin{\tab RiemannSymmetric[\{a,~b,~c,~d\}]}\\[-5pt]
	\mmanoin{]} \\
	\mmaout{\texttt{\{}g^{bd} R^{ac} -  g^{bc} R^{ad} -  g^{ad} R^{bc} + g^{ac} R^{bd} + \tfrac{1}{2} g^{ad} g^{bc} R -  \tfrac{1}{2} g^{ac} g^{bd} R -  R^{abcd}\texttt{, }}\\
	\mmanoout{\phantom{\{}g^{bd} R^{ac} -  g^{bc} R^{ad} -  g^{ad} R^{bc} + g^{ac} R^{bd} + \tfrac{1}{2} g^{ad} g^{bc} R -  \tfrac{1}{2} g^{ac} g^{bd} R - 2 R^{abcd} + R^{acbd} -  R^{adbc}\texttt{, }}\\
	\mmanoout{\phantom{\{}g^{bd} R^{ac} -  g^{bc} R^{ad} -  g^{ad} R^{bc} + g^{ac} R^{bd} + \tfrac{1}{2} g^{ad} g^{bc} R -  \tfrac{1}{2} g^{ac} g^{bd} R -  R^{acbd} + R^{adbc}\texttt{, }}\\
	\mmanoout{\phantom{\{}R^{abcd} -  R^{acbd} + R^{adbc}\texttt{\}}}
\end{mma}
Converting the above to Weyl tensors, we find:
\begin{mma}
	\mmain{\%~//~RiemannToWeyl~//~CollectTensors}\\ 
	\mmaout{\texttt{\{}- W[\triangledown ]^{abcd}\texttt{, }-2 W[\triangledown ]^{abcd} + W[\triangledown ]^{acbd} -  W[\triangledown ]^{adbc}\texttt{, }- W[\triangledown ]^{acbd} + W[\triangledown ]^{adbc}\texttt{, }}\\
	\mmanoout{\phantom{\{}W[\triangledown ]^{abcd} -  W[\triangledown ]^{acbd} + W[\triangledown ]^{adbc}\texttt{\}}}
\end{mma}%
As is obvious from this example, \func{ConstructDDIs}, like \funcref{AllContractions},
 does not take multi-term symmetries like the Bianchi identity into account.

\subsubsection{IndexConfigurations}
\label{IndexConfigurations}

\begin{usagetable}
	\usageline
		{IndexConfigurations[{\it expr}]}
		{gives a list of all independent index configurations of \textit{expr}.}
\end{usagetable}

\paragraph{Details}

The command \func{IndexConfigurations} gives all possible independent permutations of the free
indices of the input expression. A permutation of the free indices (or index configuration)
is independent when it cannot be related to another index configuration by canonicalizing.
The heavy lifting in \func{IndexConfigurations} is actually done by the \emph{SymManipulator} package
\cite{Backdahl:2013}, which can compute the right transversal of $H$ in $S_n$,
where $H$ is the symmetry group of the input expression, and $n$ the number of free indices.
A right transversal is the set of representatives of the right cosets $H / S_n$, which in turn
is in one-to-one correspondence to the set of independent index configurations.

\paragraph{Examples}

Here's one simple example of how to use \func{IndexConfigurations}:
\begin{mma}
	\mmain{IndexConfigurations[metric[a,~b]]}\\ 
	\mmaout{\texttt{\{}g^{ab}\texttt{\}}}
\end{mma}%
Because the metric is symmetric, there is only one index configuration. For two metrics we get:
\begin{mma}
	\mmain{IndexConfigurations[metric[a,~b]~metric[c,~d]]}\\ 
	\mmaout{\texttt{\{}g^{ad} g^{bc}\texttt{, }g^{ac} g^{bd}\texttt{, }g^{ab} g^{cd}\texttt{\}}}
\end{mma}%
And for three metrics:
\begin{mma}
	\mmain{IndexConfigurations[metric[a,~b]~metric[c,~d]~metric[e,~f]]}\\ 
	\mmaout{\texttt{\{}g^{af} g^{be} g^{cd}\texttt{, }g^{ae} g^{bf} g^{cd}\texttt{, }g^{af} g^{bd} g^{ce}\texttt{, }g^{ad} g^{bf} g^{ce}\texttt{, }g^{ae} g^{bd} g^{cf}\texttt{, }}\\
	\mmanoout{\phantom{\{}g^{ad} g^{be} g^{cf}\texttt{, }g^{af} g^{bc} g^{de}\texttt{, }g^{ac} g^{bf} g^{de}\texttt{, }g^{ab} g^{cf} g^{de}\texttt{, }g^{ae} g^{bc} g^{df}\texttt{, }}\\
	\mmanoout{\phantom{\{}g^{ac} g^{be} g^{df}\texttt{, }g^{ab} g^{ce} g^{df}\texttt{, }g^{ad} g^{bc} g^{ef}\texttt{, }g^{ac} g^{bd} g^{ef}\texttt{, }g^{ab} g^{cd} g^{ef}\texttt{\}}}
\end{mma}%
Lastly, for the Riemann tensor we obtain:
\begin{mma}
	\mmain{IndexConfigurations[RiemannCD[-a,~-b,~-c,~-d]]}\\ 
	\mmaout{\texttt{\{}R_{abcd}\texttt{, }R_{acbd}\texttt{, }R_{adbc}\texttt{\}}}
\end{mma}%
Note that \func{IndexConfigurations} does not take multi-term symmetries like the Bianchi identity into account, 
and hence it does not see that the last term can actually be written in terms of the first two.

\subsubsection{MakeAnsatz}
\label{MakeAnsatz}

\begin{usagetable}
	\usageline
		{MakeAnsatz[\{$e_1, e_2, \cdots $\}]}
		{returns $C_1 e_1 + C_2 e_2 + \ldots$, where the $C_i$'s are newly defined constant symbols.}
\end{usagetable}

\noindent \func{MakeAnsatz} is a convenience function that, out of a list of terms, constructs an Ansatz
with constant Symbols. Here's an example of how it works:
\begin{mma}
	\mmain{MakeAnsatz[\{metric[-a,~-b],~RicciCD[-a,~-b]\}]}\\ 
	\mmaout{C_1 g_{ab} + C_2 R_{ab}}
\end{mma}%
Even though the constant symbols print as $\tt{C_i}$, their Mathematica symbol name is $\tt{Ci}$:
\begin{mma}
	\mmain{\{C1,~C2\}}\\ 
	\mmaout{\texttt{\{}C_1^{\text{}}\texttt{, }C_2^{\text{}}\texttt{\}}}
\end{mma}%
In combination with other functions such as \funcref{IndexConfigurations} or \funcref{AllContractions}, 
\func{MakeAnsatz} becomes very handy:
\begin{mma}
	\mmain{MakeAnsatz~@~IndexConfigurations[metric[a,~b]~metric[c,~d]]}\\ 
	\mmaout{C_1 g^{ad} g^{bc} + C_2 g^{ac} g^{bd} + C_3 g^{ab} g^{cd}}
\end{mma}%
\begin{mma}
	\mmain{MakeAnsatz~@~AllContractions[}\\[-5pt]
	\mmanoin{\tab RiemannCD[a,~b,~c,~d]~RiemannCD[e,~f,~g,~h]}\\[-5pt]
	\mmanoin{]}\\ 
	\mmaout{C_1 R_{ab} R^{ab} + C_2 R^2 + C_3 R_{abcd} R^{abcd} + C_4 R_{acbd} R^{abcd}}
\end{mma}%

\subsection{Tensor algebra}

This section describes the functions in \xTras~that can be used for doing basic algebra with tensors.
There are two functions for rewriting expressions (\funcref{CollectTensors} and 
\hyperref[CollectConstants]{\func{Collect-}} \hyperref[CollectConstants]{\func{Constants}}),
and two functions for solving equations (\funcref{SolveTensors} and \funcref{SolveConstants}).

\subsubsection{CollectTensors}
\label{CollectTensors}

\begin{usagetable}
	\usageline
		{CollectTensors[{\it expr}]}
		{collects all tensorial terms in \textit{expr}.}
\end{usagetable}
\noindent \func{CollectTensors} works like the Mathematica function \func{Collect}, with the difference
that you do not have to specify a second argument: it collects all tensorial terms it can find in the
input expression. A `tensorial term' is a single tensor, or a product of tensors that 
cannot be expanded into a sum.

For example, assuming the scalars \func{T1[]}, \func{T2[]}, and \func{T3[]} are defined, 
we can make the following expression:
\begin{mma}
	\mmain{expr~=~MakeAnsatz[}\\[-5pt]
	\mmanoin{\tab \{T1[],~T1[],~T2[],~T2[],~T3[],~T3[],~T1[]~T2[],~T1[]~T2[]\}}\\[-5pt]
	\mmanoin{]}\\ 
	\mmaout{C_1 T1 + C_2 T1 + C_3 T2 + C_4 T2 + C_7 T1 T2 + C_8 T1 T2 + C_5 T3 + C_6 T3}
\end{mma}%
By calling \func{CollectTensors}, the tensors in this expression will be collected together:
\begin{mma}
	\mmain{CollectTensors[expr]}\\ 
	\mmaout{(C_1 + C_2) T1 + (C_3 + C_4) T2 + (C_7 + C_8) T1 T2 + (C_5 + C_6) T3}
\end{mma}%
\func{CollectTensors} also handles non-scalar tensors, which by default will be canonicalized
before being collected:
\begin{mma}
	\mmain{CollectTensors[}\\[-5pt]
	\mmanoin{\tab C1~RicciCD[-b,~-a]~+~C2~metric[-a,~-c]~RicciCD[c,~-b]}\\[-5pt]
	\mmanoin{]}\\ 
	\mmaout{(C_1 + C_2) R_{ab}}
\end{mma}%

\subsubsection{CollectConstants}
\label{CollectConstants}

\begin{usagetable}
	\usageline
		{CollectConstants[{\it expr}]}
		{collects all constant symbols in \textit{expr}.}
\end{usagetable}

\noindent \func{CollectConstants} is the sibling of \funcref{CollectTensors}. Instead of collecting all 
tensorial terms in the input expression, it collects all constant symbols it can find.
For example:
\begin{mma}
	\mmain{CollectConstants[}\\[-5pt]
	\mmanoin{\tab C1~T1[]~+~C1~T2[]~+~C2~T3[]~+~C2~T1[]~+~C3~T3[]~+~C3~T1[]~T2[]}\\[-5pt]
	\mmanoin{]}\\ 
	\mmaout{C_1 (T1 + T2) + C_2 (T1 + T3) + C_3 (T1 \,  T2 + T3)}
\end{mma}%

\subsubsection{SolveConstants}
\label{SolveConstants}

\begin{usagetable}
	\usagelinelong
		{SolveConstants[{\it expr}]}
		{attempts to solve the system \textit{expr} of tensorial equations for all constant symbols appearing in \textit{expr}.}
\end{usagetable}


\noindent The function \func{SolveConstants} solves equations with respect to constant symbols.
Not only does it do that, it also makes sure no tensors 
appear on the right-hand-side of the solutions. To achieve this, it uses the following three-step procedure:
\begin{enumerate}
	\item Use \func{CollectTensors} on the equation to group tensorial terms.
	\item Read of equations for the prefactors from each of the tensorial terms.
	\item Solve the prefactor equations simultaneously with built-in Mathematica function \func{Solve}.
\end{enumerate}
To illustrate this procedure, take for example the same expression we had in 
\autoref{CollectTensors}, namely
\begin{mma}
	\mmain{expr~=~MakeAnsatz[}\\[-5pt]
	\mmanoin{\tab \{T1[],~T1[],~T2[],~T2[],~T3[],~T3[],~T1[]~T2[],~T1[]~T2[]\}}\\[-5pt]
	\mmanoin{]}\\ 
	\mmaout{C_1 T1 + C_2 T1 + C_3 T2 + C_4 T2 + C_7 T1 T2 + C_8 T1 T2 + C_5 T3 + C_6 T3}
\end{mma}%
The first step towards solving the equation \func{expr == 0} for the constant symbols
$C_i$ is to collect the tensorial terms:
\begin{mma}
	\mmain{CollectTensors[expr]}\\ 
	\mmaout{(C_1 + C_2) T1 + (C_3 + C_4) T2 + (C_7 + C_8) T1 T2 + (C_5 + C_6) T3}
\end{mma}%
The second step is to read off equations for the constant symbols from each tensorial term. 
The \xTras~function \func{ToConstantSymbolEquations} does exactly this:
\begin{mma}
	\mmain{ToConstantSymbolEquations[\%~==~0]}\\ 
	\mmaout{C_1 + C_2^{\text{}}~\texttt{==}~0~\texttt{\&\&}~C_3 + C_4^{\text{}}~\texttt{==}~0~\texttt{\&\&}~C_5 + C_6^{\text{}}~\texttt{==}~0~\texttt{\&\&}~C_7 + C_8^{\text{}}~\texttt{==}~0}
\end{mma}%
The result is then fed into \func{Solve}:
\begin{mma}
	\mmain{Solve[\%,~\{C2,~C4,~C6,~C8\}]}\\ 
	\mmaout{\texttt{\{}\texttt{\{}C_2^{\text{}} \rightarrow - C_1^{\text{}}\texttt{, }C_4^{\text{}} \rightarrow - C_3^{\text{}}\texttt{, }C_6^{\text{}} \rightarrow - C_5^{\text{}}\texttt{, }C_8^{\text{}} \rightarrow - C_7^{\text{}}\texttt{\}}\texttt{\}}}
\end{mma}%
Indeed, this is the same answer we would have gotten if we had directly asked \func{SolveConstants}:
\begin{mma}
	\mmain{SolveConstants[expr~==~0]}\\ 
	\mmaout{\texttt{\{}\texttt{\{}C_2^{\text{}} \rightarrow - C_1^{\text{}}\texttt{, }C_4^{\text{}} \rightarrow - C_3^{\text{}}\texttt{, }C_6^{\text{}} \rightarrow - C_5^{\text{}}\texttt{, }C_8^{\text{}} \rightarrow - C_7^{\text{}}\texttt{\}}\texttt{\}}}
\end{mma}%


\subsubsection{SolveTensors}
\label{SolveTensors}

\begin{usagetable}
	\usageline
		{SolveTensors[{\it expr}]}
		{attempts to solve the system \textit{expr} of tensorial equations for all tensors in \textit{expr}.}
	\usageline
		{SolveTensors[{\it expr}, {\it tens}]}
		{attempts to solve the system \textit{expr} of tensorial equations for the tensors \textit{tens}.}
\end{usagetable}

\noindent Solving equations for tensors in an automated fashion is a tricky proposition. Not only does one have 
to deal with dummy indices and different forms of tensors, but also with the fact the equations
may be solved only after taking one or more contractions. \func{SolveTensors} does not address
these issues; instead, it rather solves tensorial equations for any (product of) tensor(s) that is
not contracted with another tensor. This does not always return the most general space of solutions,
but a subset of it.

For example, it solves the Einstein equation as
\begin{mma}
	\mmain{SolveTensors[}\\[-5pt]
	\mmanoin{\tab RicciCD[-a,~-b]~-~1/2~metric[-a,~-b]~RicciScalarCD[]~==~0}\\[-5pt]
	\mmanoin{]} \\
	\mmaout{\texttt{\{}\texttt{\{}\texttt{HoldPattern[}R{}^{\uunderline{a}}{}^{\uunderline{b}}\texttt{]}~\ruledelayed \texttt{Module[}\texttt{\{}\texttt{\}, } \tfrac{1}{2} g^{ba} R\texttt{]}\texttt{\}}\texttt{\}}}
\end{mma}%
The double line underneath the indices on the left-hand-side ensures that all Ricci tensors
get replaced when using this rule, regardless whether their indices are up or down:
\begin{mma}
	\mmain{RicciCD[c,~-d]~/.~\%}\\ 
	\mmaout{\texttt{\{}\tfrac{1}{2} \delta_{d}{}^{c} R\texttt{\}}}
\end{mma}%
In some simple cases, \func{SolveTensors} does return the general solution,
\begin{mma}
	\mmain{SolveTensors[T3[]~T1[]~==~T3[]~T2[]]}\\ 
	\mmaout{\texttt{\{}\texttt{\{}\texttt{HoldPattern[}T2\texttt{]}~\ruledelayed \texttt{Module[}\texttt{\{}\texttt{\}}\texttt{, }T1\texttt{]}\texttt{\}}\texttt{, }}\\
	\mmanoout{\phantom{\texttt{\{}}\texttt{\{}\texttt{HoldPattern[}T3\texttt{]}~\ruledelayed \texttt{Module[}\texttt{\{}\texttt{\}}\texttt{, }0\texttt{]}\texttt{\}}\texttt{\}}}
\end{mma}%
but, as said, in general it does not. Hence \func{SolveTensors} should more be used as a way to easily obtain 
proper \xAct~tensor replacement rules than as a method to solve generic tensorial equations.

It is worth mentioning that the second argument of \func{SolveTensors}, which specifies what
tensors to solve for, also takes patterns:
\begin{mma}
	\mmain{SolveTensors[}\\[-5pt]
	\mmanoin{\tab RicciCD[-a,~-b]~-~1/2~metric[-a,~-b]~RicciScalarCD[]~==~0,}\\[-5pt]
	\mmanoin{\tab metric[\_\_]}\\[-5pt]
	\mmanoin{]}\\ 
	\mmaout{\texttt{\{}\texttt{\{}\texttt{HoldPattern[}g{}^{\uunderline{a}}{}^{\uunderline{b}}\texttt{]}~\ruledelayed \texttt{Module[}\texttt{\{}\texttt{\}}\texttt{, }\dfrac{2 R^{ab}}{R}\texttt{]}\texttt{\}}\texttt{\}}}
\end{mma}%
Because the pattern \func{metric[\_\_]} matches the explicit form \func{~metric[-a,~-b]}, this solved
for the metric. For higher rank tensors using patterns is particularly convenients, as this avoids
having to type all indices.

\subsection{Young tableaux}
\label{sec:youngtableaux}

Conspicuously absent in \xAct~are functions that deal with Young tableaux and multi-term symmetries. 
\xTras~provides a few functions in an attempt to partly fill this void, but it is by no means a complete
treatment of the subject.

\subsubsection{YoungProject}
\label{YoungProject}

\begin{usagetable}
	\usageline
		{YoungProject[{\it expr}, {\it tab}]}
		{projects the tensorial expression \textit{expr} onto the Young tableau \textit{tab}.}
\end{usagetable}

\paragraph{Details}

If you try to antisymmetrize the Riemann tensor over three indices in \xAct, you will find that
the result is non-zero:
\begin{mma}
	\mmain{ToCanonical~@~Antisymmetrize[RiemannCD[-a,-b,-c,-d],~\{-a,-b,-c\}]}\\ 
	\mmaout{\tfrac{1}{3} R_{abcd} -  \tfrac{1}{3} R_{acbd} + \tfrac{1}{3} R_{adbc}}
\end{mma}%
This is because \func{ToCanonical} does not take multi-term symmetries, like the Bianchi identity
$R_{[abc]d} = 0$, into account. However, these symmetries can be made explicit by projecting tensors
onto their respective Young tableaux \cite{Green:2005qr}. The projection can be done with so-called
Young projectors \cite{Etingof:2009ab}, which are sequential row-by-row symmetrizations and
column-by-column antisymmetrizations of the Young tableau. To be precise, if we have a Young diagram
$\lambda$ (i.e. a partition of the integer $n$) and one of its Young tableaux $\lambda_a$, then 
the Young projector reads
\begin{equation}
	P_A^{\lambda_a} = \frac{ f^\lambda}{n!} 
		\prod_{k \in \textrm{col}(\lambda_a)} A^k 
		\prod_{l \in \textrm{row}(\lambda_a)} S^l 
\end{equation}
where $f^\lambda$ is the dimension of the Young diagram, 
and $S^n$ ($A^n$) the (anti-)symmetrization of the $n^\textrm{nth}$
row (column). Here, both $S$ and $A$ are without any weight, i.e.~$S ( \{x,y\} ) = \{x,y\} + \{y,x\}$
and not $\frac{1}{2}(\{x,y\} + \{y,x\})$.

The above Young projector is manifestly antisymmetric, because the columns are antisymmetrized
after the rows are symmetrized. Changing this order gives the manifestly symmetric Young projector
$P_S$:
\begin{equation}
	P_S^{\lambda_a} = \frac{ f^\lambda}{n!} 
		\prod_{l \in \textrm{row}(\lambda_a)} S^l 
		\prod_{k \in \textrm{col}(\lambda_a)} A^k 		
\end{equation}
which is also a perfectly fine projector. By default, \func{YoungProject} uses the manifestly
antisymmetric projector $P_A$, but by setting the option \func{ManifestSymmetry} to \func{Symmetric}
it is possible to use the manifestly symmetric projector $P_S$.

\paragraph{Examples}

\ytableausetup{smalltableaux,centertableaux}

Projecting a tensor $S^{ab}$ onto the Young tableau~\begin{ytableau}a & b\end{ytableau}~can be
done as follows:
\begin{mma}
	\mmain{YoungProject[S[a,~b],~\{\{a,~b\}\}]}\\ 
	\mmaout{\tfrac{1}{2} S^{ab} + \tfrac{1}{2} S^{ba}}
\end{mma}%
And projecting it onto the tableau \begin{ytableau}a \\ b\end{ytableau} gives:
\begin{mma}
	\mmain{YoungProject[S[a,~b],~\{\{a\},~\{b\}\}]}\\ 
	\mmaout{\tfrac{1}{2} S^{ab} -  \tfrac{1}{2} S^{ba}}
\end{mma}%
Projecting the Riemann tensor onto the tableau \begin{ytableau}a & c\\ b & d \end{ytableau}
goes as follows:
\begin{mma}
	\mmain{YoungProject[RiemannCD[-a,~-b,~-c,~-d],~\{\{-a,~-c\},~\{-b,~-d\}\}]}\\ 
	\mmaout{\tfrac{2}{3} R_{abcd} + \tfrac{1}{3} R_{acbd} -  \tfrac{1}{3} R_{adbc}}
\end{mma}%
And indeed, the Bianchi identity is manifest after projection:
\begin{mma}
	\mmain{ToCanonical~@~Antisymmetrize[\%,~\{-a,~-b,~-c\}]}\\ 
	\mmaout{0}
\end{mma}%

By default, \func{YoungProject} uses a manifestly antisymmetric projection. It projects for example
a rank-3 tensor $T^{abc}$ onto the Young tableau \begin{ytableau}a & b\\ c  \end{ytableau} as
\begin{mma}
	\mmain{YoungProject[T[a,~b,~c],~\{\{a,~b\},~\{c\}\}]}\\ 
	\mmaout{\tfrac{1}{3} T^{abc} + \tfrac{1}{3} T^{bac} -  \tfrac{1}{3} T^{bca} -  \tfrac{1}{3} T^{cba}}
\end{mma}%
which is indeed antisymmetric in $a$ and $c$. We can switch to a manifestly symmetric
projection with the option \func{ManifestSymmetry}:
\begin{mma}
	\mmain{YoungProject[}\\[-5pt]
	\mmanoin{\tab T[a,~b,~c],}\\[-5pt]
	\mmanoin{\tab \{\{a,~b\},~\{c\}\},}\\[-5pt]
	\mmanoin{\tab ManifestSymmetry~->~Symmetric}\\[-5pt]
	\mmanoin{]}\\ 
	\mmaout{\tfrac{1}{3} T^{abc} + \tfrac{1}{3} T^{bac} -  \tfrac{1}{3} T^{cab} -  \tfrac{1}{3} T^{cba}}
\end{mma}%
The result is now no longer antisymmetric in $a$ and $c$, but symmetric in $a$ and $b$.

\subsubsection{RiemannYoungProject}
\label{RiemannYoungProject}

\ytableausetup{smalltableaux,aligntableaux=bottom}

\begin{usagetable}
	\usagelinelong
		{RiemannYoungProject[{\it expr}]}
		{projects all Riemann tensors and their first derivatives in \textit{expr} onto their Young tableaux. }
\end{usagetable}

\noindent The function \func{RiemannYoungProject} automatizes the projection of Riemann tensors onto their Young tableaux;
it replaces every occurrence of a Riemann tensor or a first derivative of it with their Young
projected versions. That is, it does the replacements
\begin{subequations}
\begin{align}
	R_{abcd} \rightarrow & \; P_A^{\textrm{\tiny\begin{ytableau}a & c\\ b & d \end{ytableau}}} \left( R_{abcd} \right) , \\
	\nabla_e R_{abcd} \rightarrow & \; P_A^{\textrm{\tiny\begin{ytableau}a & c & e \\ b & d \end{ytableau}}} \left(\nabla_e R_{abcd} \right) .
\end{align}
\end{subequations}
For example, a single Riemann tensor is replaced as follows:
\begin{mma}
	\mmain{RiemannYoungProject~@~RiemannCD[-a,~-b,~-c,~-d]}\\ 
	\mmaout{\tfrac{2}{3} R_{abcd} + \tfrac{1}{3} R_{acbd} -  \tfrac{1}{3} R_{adbc}}
\end{mma}%
A first derivative of the Riemann tensor gets replaced as:
\begin{mma}
	\mmain{RiemannYoungProject[CD[-e]~@~RiemannCD[-a,~-b,~-c,~-d]]}\\ 
	\mmaout{\tfrac{1}{12} \triangledown_{a}R_{bcde} -  \tfrac{1}{12} \triangledown_{a}R_{bdce} -  \tfrac{1}{6} \triangledown_{a}R_{becd} -  \tfrac{1}{12} \triangledown_{b}R_{acde} + \tfrac{1}{12} \triangledown_{b}R_{adce} + \tfrac{1}{6} \triangledown_{b}R_{aecd}}\\ 
	\mmanoout{ -  \tfrac{1}{6} \triangledown_{c}R_{abde} -  \tfrac{1}{12} \triangledown_{c}R_{adbe} + \tfrac{1}{12} \triangledown_{c}R_{aebd} + \tfrac{1}{6} \triangledown_{d}R_{abce} + \tfrac{1}{12} \triangledown_{d}R_{acbe} -  \tfrac{1}{12} \triangledown_{d}R_{aebc}}\\ 
	\mmanoout{ + \tfrac{1}{3} \triangledown_{e}R_{abcd} + \tfrac{1}{6} \triangledown_{e}R_{acbd} -  \tfrac{1}{6} \triangledown_{e}R_{adbc}}
\end{mma}%
This enables us to easily prove e.g.~the second Bianchi identity $\nabla_{[a} R_{bc]de}=0$:
\begin{mma}
	\mmain{ToCanonical~@~RiemannYoungProject~@~Antisymmetrize[}\\[-5pt]
	\mmanoin{\tab CD[-a]~@~RiemannCD[-b,~-c,~-d,~-e],}\\[-5pt]
	\mmanoin{\tab \{-a,~-b,~-c\}}\\[-5pt]
	\mmanoin{]}\\ 
	\mmaout{0}
\end{mma}%
Another nice example is the identity $R_{acde} R^{bdce} = \frac{1}{2} R_{acde} R^{bcde}$, which
can be proven as follows:
\begin{mma}
	\mmain{ToCanonical @ RiemannYoungProject[}\\[-5pt]
	\mmanoin{\tab RiemannCD[-a,-c,-d,-e](RiemannCD[b,d,c,e]-1/2RiemannCD[b,c,d,e])}\\[-5pt]
	\mmanoin{]}\\ 
	\mmaout{0}
\end{mma}%

\subsubsection{TableauSymmetric}
\label{TableauSymmetric}

\ytableausetup{smalltableaux,centertableaux}

\begin{usagetable}
	\usageline
		{TableauSymmetric[{\it tab}]}
		{gives the symmetry of the tableau \textit{tab}.}
\end{usagetable}

\noindent \func{TableauSymmetric} generalizes the \xAct~functions \func{Symmetric}, \func{Antisymmetric},
and \func{Riemann-} \func{Symmetric} to arbitrary Young tableaux. This comes in particularly handy
when defining tensors that have more complicated symmetry structures than just complete
(anti-)symmetry. Say, for instance, we 
have a tensor $T^{abcdef}$ that lives in the Young diagram {\tiny{\ydiagram{3,2,1}}}. If we
define it without any symmetry,
\begin{mma}
	\mmain{DefTensor[T[a,b,c,d,e,f], M]}
\end{mma}
and subsequently project it onto its Young tableau, we get no less than 144 terms:
\begin{mma}
	\mmain{Length~@~YoungProject[}\\[-5pt]
	\mmanoin{\tab T[a,~b,~c,~d,~e,~f],}\\[-5pt]
	\mmanoin{\tab \{\{a,~b,~c\},~\{d,~e\},~\{f\}\}}\\[-5pt]
	\mmanoin{]}\\ 
	\mmaout{144}
\end{mma}%
However, if we had instead defined it with the appropriate symmetry,
\begin{mma}
	\mmain{DefTensor[}\\[-5pt]
	\mmanoin{\tab T[a,b,c,d,e,f], M,}\\[-5pt]
	\mmanoin{\tab TableauSymmetric[\{\{a,b,c\},~\{d,e\},~\{f\}\}}\\[-5pt]
	\mmanoin{]}
\end{mma}
we would have gotten just 57 terms:
\begin{mma}
	\mmain{Length~@~YoungProject[}\\[-5pt]
	\mmanoin{\tab T[a,~b,~c,~d,~e,~f],}\\[-5pt]
	\mmanoin{\tab \{\{a,~b,~c\},~\{d,~e\},~\{f\}\}}\\[-5pt]
	\mmanoin{]}\\ 
	\mmaout{57}
\end{mma}%
This is because the tensor $T^{abcdef}$ now has all the mono-term symmetries that come
from its Young diagram. For example,
\begin{mma}
	\mmain{ToCanonical[T[f,~e,~c,~a,~b,~d]]}\\ 
	\mmaout{- T^{abcdef}}
\end{mma}%
It are these mono-term symmetries that reduce the number of terms in the Young projection.

\subsection{Miscellaneous}

Lastly, this section describes some \xTras~functions that do not fall in any of the other categories.

\subsubsection{VarL}
\label{VarL}

\begin{usagetable}
	\usageline
		{VarD[{\it g}[-{\it a},-{\it b}], {\it cd}][{\it S}]}
		{returns $\tfrac{\delta S}{\delta g_{ab}}$ while integrating by parts with respect to the covariant derivative \textit{cd}.}
	\usagelinelong
		{VarL[{\it g}[-{\it a},-{\it b}], {\it cd}][{\it L}]}
		{returns $\tfrac{1}{\sqrt{\lvert g \rvert}}\tfrac{\delta \sqrt{\lvert g \rvert} L}{\delta g_{ab}}$ while integrating by parts with respect to the covariant derivative \textit{cd}.}
\end{usagetable}

\paragraph{Details} 

Because of the non-linear metric dependence of curvature tensors, computing their equations of motion
with respect to the metric can be a rather involved affair. While the variation of the Einstein-Hilbert
term is relatively easy, things like
\begin{equation}
	\frac{\delta}{\delta g_{ab}} 
	\left( 
		R_{cd}{}^{gh} R^{cdef} \nabla_{f}\nabla_{l}R_{h}{}^{l}{}_{ik} \nabla_{j}\nabla_{g}R_{e}{}^{ijk}	\right)
\end{equation}
can be quite cumbersome. By using the power of the \emph{xPert} package \cite{Brizuela:2008ra}, 
\xTras~can compute variations like the above with relative ease. It does this by first computing 
the total variation, and then integrating by parts. Schematically, this reads
\begin{equation}
	\delta F
		 = f_1 \delta g + f_2 \nabla \delta g + 
			f_3 \nabla \nabla \delta g + \cdots
		 = \frac{\delta F}{\delta g} \delta g + \textrm{total derivative}
\end{equation}
where $F$ and $f_i$ are a functionals that depend on the metric $g$, and $\frac{\delta F}{\delta g}$
is the quantity we're after.

Computing the total variation is the first step towards reading off $\frac{\delta F}{\delta g}$, 
and is carried out by the \emph{xPert} commands \func{Perturbation} and \func{ExpandPerturbation}:
\begin{mma}
	\mmain{ExpandPerturbation~@~Perturbation[RicciScalarCD[]]}\\[-5pt]
	\mmanoin{\tab//~ContractMetric~//~ToCanonical}\\ 
	\mmaout{- \triangle g^{1ab} R_{ab} + \triangledown_{b}\triangledown_{a}\triangle g^{1ab} -  \triangledown_{b}\triangledown^{b}\triangle g^{1a}{}_{a}}
\end{mma}%
Here $\tt{\triangle g^{1}_{ab}}$ is the same as $\delta g_{ab}$ above, namely the perturbation of the metric.
The second step, integrating by parts and peeling off $\delta g_{ab}$, is done with the \emph{xTensor}
command \func{VarD}:
\begin{mma}
	\mmain{VarD[$\tt{\triangle g^{1}_{ab}}$, CD][\%]}\\ 
	\mmaout{- \delta_{1}{}^{1} g^{ac} g^{bd} R_{cd}}
\end{mma}%
The spurious $\tt{\delta_{1}{}^{1}}$ comes from the way \xAct~handles the variation 
$\frac{\delta \tt{\triangle g^{1}_{ab}}}{\delta \tt{\triangle g^{1}_{cd}}}$ and is equal to one, 
even though it is not automatically simplified.

\xTras~overwrites the \func{VarD} command such that the above two-step procedure is carried
out whenever the variation is with respect to a metric. When the variation is with respect
to another tensor, \emph{xTensor}'s \func{VarD} is used.

\paragraph{Examples}

The variation of the Ricci scalar with respect to the metric can be computed with the following command:
\begin{mma}
	\mmain{VarD[metric[-a,~-b],~CD][RicciScalarCD[]]}\\ 
	\mmaout{- g^{ac} g^{bd} R_{cd}}
\end{mma}%
Using \func{VarL} instead of \func{VarD} automatically takes care of overall factors of $\sqrt{\lvert g \rvert}$:
\begin{mma}
	\mmain{VarL[metric[-a,~-b],~CD][RicciScalarCD[]]}\\ 
	\mmaout{- g^{ac} g^{bd} R_{cd} + \tfrac{1}{2} g^{ab} R}
\end{mma}%
Note that \func{VarD} and \func{VarL} do not contract metrics and canonicalize on their own. If we want, we have to do
this ourselves afterwards.
Varying the Einstein-Hilbert term coupled to a scalar field $\phi$ with respect to the metric gives:
\begin{mma}
	\mmain{VarL[metric[-a,~-b],~CD][phi[]~RicciScalarCD[]]}\\[-5pt]
	\mmanoin{\tab //~ContractMetric~//~ToCanonical}\\ 
	\mmaout{- \phi R^{ab} + \tfrac{1}{2} g^{ab} \phi R + \tfrac{1}{2} \triangledown^{a}\triangledown^{b}\phi + \tfrac{1}{2} \triangledown^{b}\triangledown^{a}\phi -  g^{ab} \triangledown_{c}\triangledown^{c}\phi}
\end{mma}%
Higher powers of $R$ can also be varied easily:
\begin{mma}
	\mmain{VarL[metric[-a,~-b],~CD][RicciScalarCD[]\textasciicircum 2]}\\[-5pt]
	\mmanoin{\tab //~ContractMetric~//~ToCanonical}\\ 
	\mmaout{-2 R^{ab} R + \tfrac{1}{2} g^{ab} R^2 + \triangledown^{a}\triangledown^{b}R + \triangledown^{b}\triangledown^{a}R - 2 g^{ab} \triangledown_{c}\triangledown^{c}R}
\end{mma}%
And higher still:
\begin{mma}
	\mmain{VarL[metric[-a,~-b],~CD][RicciScalarCD[]\textasciicircum 4]}\\[-5pt]
	\mmanoin{\tab //~ContractMetric~//~ToCanonical}\\ 
	\mmaout{-4 R^{ab} R^3 + \tfrac{1}{2} g^{ab} R^4 + 6 R^2 \triangledown^{a}\triangledown^{b}R + 24 R \triangledown^{a}R \triangledown^{b}R + 6 R^2 \triangledown^{b}\triangledown^{a}R - 12 g^{ab} R^2 \triangledown_{c}\triangledown^{c}R}\\ 
	\mmanoout{ - 24 g^{ab} R \triangledown_{c}R \triangledown^{c}R}
\end{mma}%

\subsubsection{FullSimplification}
\label{FullSimplification}

\begin{usagetable}
	\usagelinelong
		{FullSimplification[][{\it expr}]}
		{tries to simplify \textit{expr} as much as possible, taking Bianchi identities into account and sorting covariant derivatives.}
\end{usagetable}

\noindent When dealing with curvature tensors, it is often desirable to use the Bianchi identities 
to rewrite expression in the simplest form possible. \func{ToCanonical} cannot be used for this,
since it only simplifies mono-term symmetries, and Bianchi identities are multi-term symmetries.
The Bianchi identities are however implemented in the simplification methods of the \emph{Invar} package 
\cite{MartinGarcia:2007ab,MartinGarcia:2008qz}. But unfortunately, \emph{Invar} can only simplify
scalar monomials of Riemann tensors.

The function \func{FullSimplification} extends the capabilities of \emph{Invar} slightly by
also simplifying the contracted second Bianchi identities in any expression, not just scalar monomials.
When given an input expression, \func{FullSimplification} does the following:
\begin{enumerate}
	\item Simplify scalar monomials with the help of the \emph{Invar} package.
	\item Apply the contracted second Bianchi identities $\nabla_a R_{bcd}{}^a = \nabla_c R_{bd} - \nabla_b R_{cd}$
	and $\nabla_a R_b{}^a = \tfrac 1 2 \nabla_b R$.
	\item Sort covariant derivatives.
\end{enumerate}
For example, when given the expression $\nabla^a \nabla_b R_{ca}$, \func{FullSimplification} commutes
covariant derivatives to divergences such that it can use the contracted Bianchi identities, and then 
afterwards sorts covariant derivatives:
\begin{mma}
	\mmain{FullSimplification[][CD[a]~@~CD[-b]~@~RicciCD[-c,~-a]]}\\ 
	\mmaout{R_{b}{}^{a} R_{ca} -  R^{ad} R_{bacd} + \tfrac{1}{2} \triangledown_{c}\triangledown_{b}R}
\end{mma}%
Note that covariant derivatives are sorted with the \xAct~command \func{SortCovDs}, which sorts
them in alphabetical order in postfix notation. Thus $\nabla_c \nabla_b R = R_{;b;c}$ is sorted.

As said, \func{FullSimplification} also simplifies scalar monomials by using all Bianchi identities,
not just the contracted Bianchi identities:
\begin{mma}
	\mmain{FullSimplification[][RiemannCD[a,b,c,d]~RiemannCD[-a,-c,-b,-d]]}\\ 
	\mmaout{\tfrac{1}{2} R_{abcd} R^{abcd}}
\end{mma}%
This is a contraction of the identity we found in \autoref{RiemannYoungProject}.

\subsubsection{EulerDensity}
\label{EulerDensity}

\begin{usagetable}
	\usageline
		{EulerDensity[{\it cd}]}
		{gives the Euler density associated to the covariant derivative \textit{cd}.}
	\usagelinelong
		{EulerDensity[{\it cd}, {\it dim}]}
		{gives the Euler density associated to the covariant derivative \textit{cd} in the dimension
		\textit{dim} if the underlying manifold has a generic dimension.}
\end{usagetable}

\paragraph{Details}

The Euler density $E_{2n}$ in dimension $d = 2n$ is given by
\begin{equation}
	E_{2n} = 
	\frac{1}{2^n} R_{i_1 i_2 j_1 j_2} \cdots R_{i_{n-1}i_n j_{n-1}j_n} 
	\epsilon^{i_1 \cdots i_n} \epsilon^{j_1 \cdots j_n}
\end{equation}
where $\epsilon$ is the Levi-Civita tensor, not the Levi-Civita symbol. Note that this
technically is not a density because it has zero weight. In order to obtain a density, we
would need to multiply it with $\sqrt{\lvert g \rvert}$.

In order to prevent dummy index collisions, the results of \func{EulerDensity} are wrapped
in a special head \func{Scalar}, which is indicated by a red bracket. The \func{Scalar} heads can 
be removed with the \emph{xTensor} command \func{NoScalar}.

\paragraph{Examples}

\newcommand{\biglr}{\textcolor{red}{\bigl(}}
\newcommand{\bigrr}{\textcolor{red}{\bigr)}}

Because we have a manifold with generic dimension, we need to specify the second argument of
\func{EulerDensity}. For two dimensions, the Euler density reads:
\begin{mma}
	\mmain{EulerDensity[CD,~2]}\\ 
	\mmaout{- R}
\end{mma}%
And for four dimensions it is:
\begin{mma}
	\mmain{EulerDensity[CD,~4]}\\ 
	\mmaout{- R^2 + 4 \biglr R_{ab} R^{ab} \bigrr -  \biglr R_{abcd} R^{abcd} \bigrr}
\end{mma}%
In six dimensions the Euler density becomes:
\begin{mma}
	\mmain{EulerDensity[CD,~6]}\\ 
	\mmaout{- R^3 + 12 R \biglr R_{ab} R^{ab}\bigrr - 16 \biglr R_{a}{}^{c} R^{ab} R_{bc}\bigrr - 24 \biglr R^{ab} R^{cd} R_{acbd}\bigrr - 3 R \biglr R_{abcd} R^{abcd}\bigrr }\\
	\mmanoout{+ 24 \biglr R^{ab} R_{a}{}^{cde} R_{bcde}\bigrr + 8 \biglr R_{a}{}^{e}{}_{c}{}^{f} R^{abcd} R_{bfde}\bigrr - 2 \biglr R_{ab}{}^{ef} R^{abcd} R_{cdef}\bigrr}
\end{mma}%
And lastly, in eight dimensions, it is:
\begin{mma}
	\mmain{EulerDensity[CD,~8]}\\ 
	\mmaout{- R^4 + 24 R^2 \biglr R_{ab} R^{ab}\bigrr - 64 R \biglr R_{a}{}^{c} R^{ab} R_{bc}\bigrr + 96 \biglr R_{a}{}^{c} R^{ab} R_{b}{}^{d} R_{cd}\bigrr - 48 \biglr R_{ab} R^{ab}\bigrr \biglr R_{cd} R^{cd}\bigrr}\\ 
	\mmanoout{  - 96 R \biglr R^{ab} R^{cd} R_{acbd}\bigrr - 6 R^2 \biglr R_{abcd} R^{abcd}\bigrr + 96 R \biglr R^{ab} R_{a}{}^{cde} R_{bcde}\bigrr+ 384 \biglr R_{a}{}^{c} R^{ab} R^{de} R_{bdce}\bigrr }\\ 
	\mmanoout{ - 96 \biglr R^{ab} R^{cd} R_{ac}{}^{ef} R_{bdef}\bigrr - 192 \biglr R^{ab} R^{cd} R_{a}{}^{e}{}_{c}{}^{f} R_{bedf}\bigrr + 32 R \biglr R_{a}{}^{e}{}_{c}{}^{f} R^{abcd} R_{bfde}\bigrr  }\\ 
	\mmanoout{ - 8 R \biglr R_{ab}{}^{ef} R^{abcd} R_{cdef}\bigrr- 192 \biglr R_{a}{}^{c} R^{ab} R_{b}{}^{def} R_{cdef}\bigrr + 192 \biglr R^{ab} R^{cd} R_{a}{}^{e}{}_{b}{}^{f} R_{cedf}\bigrr }\\ 
	\mmanoout{- 384 \biglr R^{ab} R_{a}{}^{cde} R_{b}{}^{f}{}_{d}{}^{g} R_{cgef}\bigrr + 24 \biglr R_{ab} R^{ab}\bigrr \biglr R_{cdef} R^{cdef}\bigrr + 96 \biglr R^{ab} R_{a}{}^{cde} R_{bc}{}^{fg} R_{defg}\bigrr}\\ 
	\mmanoout{ - 192 \biglr R^{ab} R_{a}{}^{c}{}_{b}{}^{d} R_{c}{}^{efg} R_{defg}\bigrr + 96 \biglr R_{a}{}^{e}{}_{c}{}^{f} R^{abcd} R_{b}{}^{g}{}_{e}{}^{h} R_{dgfh}\bigrr + 96 \biglr R_{ab}{}^{ef} R^{abcd} R_{c}{}^{g}{}_{e}{}^{h} R_{dhfg}\bigrr}\\ 
	\mmanoout{ - 6 \biglr R_{ab}{}^{ef} R^{abcd} R_{cd}{}^{gh} R_{efgh}\bigrr + 48 \biglr R_{abc}{}^{e} R^{abcd} R_{d}{}^{fgh} R_{efgh}\bigrr - 48 \biglr R_{a}{}^{e}{}_{c}{}^{f} R^{abcd} R_{b}{}^{g}{}_{d}{}^{h} R_{egfh}\bigrr }\\ 
	\mmanoout{ - 3 \biglr R_{abcd} R^{abcd}\bigrr \biglr R_{efgh} R^{efgh}\bigrr}
\end{mma}%

\section*{Acknowledgements}

I would like to thank Thomas B\"ackdahl, Jos\'e M. Mart\'in-Garc\'ia, and Leo Stein for 
useful discussions and their contributions to the \xAct~mailing list, and Cyril Pitrou for
his suggestion to use \emph{xPert} for computing metric variations.
Furthermore, I thank Andrea Campoleoni, Massimo Taronna, and Pan Kessel for their feedback
on initial versions of \xTras.

\begingroup
\setlength{\emergencystretch}{8em}
\printbibliography
\endgroup

\end{document}